# Additively manufacturable high-strength aluminum alloys with thermally stable microstructures enabled by hybrid machine learning-based design

S. Mohadeseh Taheri-Mousavi[1,2,3,4*], Michael Xu[2], Florian Hengsbach[5], Clay Houser[2,6], Zhaoxuan Ge[3], Benjamin Glaser[3], Shaolou Wei[2], Mirko Schaper[5], James M. LeBeau[2], Greg B. Olson[2], A. John Hart[1*]

[1]Department of Mechanical Engineering, Massachusetts Institute of Technology, 77 Massachusetts Avenue, Cambridge, MA 02139, USA.
[2]Department of Materials Science and Engineering, Massachusetts Institute of Technology, 77 Massachusetts Avenue, Cambridge, MA 02139, USA.
[3]Department of Materials Science and Engineering, Carnegie Mellon University, 5000 Forbes Avenue, Pittsburgh, PA 15213, USA.
[4]Department of Mechanical Engineering, Carnegie Mellon University, 5000 Forbes Avenue, Pittsburgh, PA 15213, USA.
[5]Department of Mechanical Engineering, Paderborn University, Mersinweg 9, 33100 Paderborn, Germany
[6]Department of Materials Science and Engineering, Northwestern University, Evanston, IL 60208, USA

*Corresponding authors: smtaherimousavi@cmu.edu; ajhart@mit.edu

## Summary paragraph

Additively manufactured (AM) aluminum alloys with high strength and thermal stability have broad applications in turbine engines, vacuum pumps, heat exchangers, and many other industrial systems. Employing precipitates with an $L1_2$ structure to block dislocation motions is a widespread strategy to strengthen aluminum. However, to achieve high strength, a high volume fraction of small precipitates is required, and these characteristics are generally mutually exclusive. Here, we show that for certain compositions of Al alloys, $L1_2$ phases initially precipitate as sub-micron metastable ternary phases under the rapid solidification conditions of powder bed AM, yet the subsequent $L1_2$ phases that precipitate during heat treatment of the sample remain at the nanoscale, imparting high strength. For strength to be retained at elevated temperature, these nanoprecipitates must have low coarsening rates. To inversely design the composition of an alloy to have these target microstructural features, we used hybrid calculation of phase diagram (CALPHAD)-based integrated computational materials engineering (ICME) and Bayesian optimization techniques. We tested our approach by designing an Al-Er-Zr-Y-Yb-Ni model alloy, and the selected composition was manufactured in powder form as AM feedstock. The strength of specimens manufactured via laser powder bed fusion (LPBF) from the designed composition is comparable to that of wrought Al 7075, yet without cracking that occurs upon LPBF of Al 7075. After high-temperature (400°C) aging the designed alloy is 50% stronger than the strongest known benchmark printable Al alloy[1]. This stable strengthening strategy is applicable to a wide range of alloys and rapid solidification processes, and our hybrid ML/CALPHAD numerical framework can be used for the efficient and robust design of alloy microstructures and properties, expanding the capabilities of additive as well as traditional manufacturing.

## Introduction

Additively manufactured (AM) structural components with complex geometries and tailored properties at voxel-size resolution will lead to significant leap in performance in various critical engineering applications. One of the exceptional benefits of AM is that exploiting non-equilibrium processing pathways, including through rapid solidification, can give rise to precipitation of metastable phases[1–4], supersaturation of solutes[5–7], refining length scales of microstructural features such as grain and phase size[8–13]. These microstructural features are difficult or impossible to develop in slow solidification conditions such as during casting. For example, metastable phases due to rapid solidification have been reported in Ni, Fe, Ti, Al alloys and alter their mechanical properties or printability at different length scales[2,3,14–20]. The features may also appear in welding, giga casting, and splat quenching when the processing conditions lead to rapid solidification. Exploiting the rapid solidification may thus unravel new routes for alloy design.

Al alloys are the second widest used alloys after steels in industrial applications[21]. High (temperature) strength Al alloys have been mainly strengthened with two mechanisms[8]. Employing precipitates especially with $L1_2$ structures is the first main strategy[22]. $L1_2$ structures are preferred due to having coherent interfaces to resist coarsening at high temperatures. As the second strategy, Al grains can be partitioned with eutectic phases to further block dislocation motions [8]. Following the first strategy, strength has an inverse relationship with the precipitate size based on the Orowan strengthening theory[22]. Thus, for a given volume fraction of precipitates, smaller precipitate sizes more effectively block dislocation motions and lead to higher strength.

Although laser AM processing generally leads to smaller length scales as phases do not have time to grow due to rapid solidification, precipitate sizes usually grow to the micron scale especially as the percentage of a desirable precipitate phases increases. Another challenge is that most widely used high strength Al alloys such as Al7075 cannot be printed as they are prone to hot cracking upon rapid solidification of printed samples. [23,24]. Finally, to realize Al alloys with thermal stability (typically at 250-400°C)[8,14,25,26], it is typically necessary to add heavy high melting point elements which results in macrosegregation during casting. Because layer-wise melting of laser AM circumvents this challenge, many efforts focused on designing printable Al alloys which have strength comparable to wrought Al7075 and with high temperature stability[8,14,25,26].

Here, we show that for tailored compositions of Al alloys, rapid solidification can cause precipitation of metastable phases at submicron scales (**Fig. 1**). The subsequent heat treatment of printed samples leads to transformation of ternary phases into precipitates with $L1_2$ structures which remain at the nanoscale even at high volume fractions. We demonstrated this approach by performing a full alloy design cycle of a model Al alloy (Al-Er-Zr-Y-Yb-Ni) for AM, spanning from virtual predictions to experimental validation. The fine distribution of metastable ternary phases (Al-Ni-M, M= Er, Zr, Y, Yb) (**Fig. 1**) through rapid solidification serve as the source for the reactive elements enabling nanoscale precipitation of a high phase fraction of the thermally stable $L1_2$ ($Al_3M$ (M= Er, Zr, Y, Yb)) strengthening phases. To enable thermal stability of the microstructure, we minimized coarsening rates of $L1_2$ phases and the percentage of fast-growing phases (**Fig. 2a & Supplementary Fig. S1**). To inhibit hot cracking in the sample, we controlled the temperature range and solidification time of our alloy. As the dimensions of the compositional space (here 5 dimensions) and the number of objectives increase, defining the critical parameters governing the microstructural features and, ultimately, our desired engineering properties become significantly challenging (**Supplementary Fig. S1**).

**Results and discussion**

We used hybrid ML/CALPHAD simulations for the efficient exploration of the design space and find nonlinear correlations between design space parameters (here elemental concentrations) (**Fig. 2b**)[27–47]. Our hardness surpasses those of the best-known benchmark Al alloys which was not designed via ML techniques (**Fig. 2b**). To avoid the bottleneck of alloy design cycle (custom powder processing), we initially arc melted and laser scanned promising candidates.

Our designed alloy has 50% higher strength than our benchmark alloy which was prepared in the same way (**Fig. 1**). The microstructure and properties of two control alloys are also demonstrated to elucidate the design idea. As seen in **Fig. 1**, slower solidification via induction melting (cast) leads to micro scale ternary or/and $L1_2$ phases (**Supplementary Fig. S15**) and five times lower hardness (**Figs. 1**). We also measured the hardness and analyzed the microstructure of an Al alloy with identical molar percentage of Er and Zr but without Ni to directly precipitate $L1_2$ phases during rapid solidification and thus inhibit precipitation of ternary phases (**Fig. 1 & Supplementary Fig. S12**). SEM-energy dispersive spectroscopy (EDS) characterization shows micro scale Zr rich phases in the grain interiors (**Supplementary Fig. S16**). The hardness of this sample is 34% lower than the designed composition containing nanoscale ternary phases processed in identical procedure (**Fig. 1**, **Supplementary Fig. S12**). We then made 3D printed samples from custom powder and performed hardness and tensile tests. Our hardness is once more 50% higher than benchmark alloys at 8 hrs of aging at 400°C and comparable to wrought Al7075 (**Supplementary Fig. S17**). We further confirmed our high strength and the softening resistance of our designed alloy at elevated temperatures by performing room temperature tensile tests on as-built and aged samples (**Fig. 3**). The peak tensile strength after 8 hrs of aging is once more 50% higher than the benchmark printable Al alloy[1], and does not drop even after 48 hrs aging at 400°C. Each step in the design procedure is provided as follows below.

**Step I: Extracting microstructural features**

For our target high temperature strengthening and printability, the associated microstructural features controlling these properties are extracted as follows:

During solidification of our Al alloy system (Al-Ni-Zr-Er-Y-Yb), several precipitates form including $Al_3Zr$, $L1_2$, $Al_3Ni$, and metastable ternary phase $Al_{23}Ni_6M_4$. $Al_3Zr$ is brittle and undermines strengthening[1]; its lower solvus temperature and phase fraction minimize this phase. These quantities are microstructural features that can be quantified from single equilibrium calculations (**Figs. 5d** & **e**). $L1_2$ is stable at high-temperatures and contributes to strengthening by impeding dislocations (**Supplementary Fig. S3**)[22]. At low temperatures, based on the Orowan strengthening mechanism, dislocations bow around precipitates and leave behind dislocation loops. Based on Eq. 1 (see Methods, $\sigma_{\mathrm{or}} \propto \frac{1}{\mathrm{R}}$), the shear stress required for these mechanisms is inversely proportional to the precipitate size R, above the shear to bypass critical size of ~2 nm.

The goal is to achieve such a nanoscale dispersion for aged samples. A high at. % of $L1_2$ phase at 250°C without formation during solidification (low $L1_2$ at. % at 660°C, which is the melting temperature of pure Al) maximize this phase at small radii. These percentages are calculated from single equilibrium calculations (**Figs. 5b** & **c**). It is noteworthy that metastable precipitates such as $Al_{23}Ni_6M_4$, in which M can potentially be Er, Zr, Y, or Yb, may form at small scales during rapid solidification, working as efficient reactant reservoirs for the $L1_2$ precipitates during aging. The volume fractions of these precipitates after solidification must be maximized and are calculated from non-equilibrium Scheil solidification simulations (**Fig. 5f**).

To maintain strength at higher temperatures, thermal stability of the microstructure is required. Thus, the coarsening of $L1_2$ precipitates and the volume fraction of rapidly coarsening $Al_3Ni$ must be minimized (**Figs. 5a** & **f**). Based on Eq. 2 in the Methods Section ($k \propto \frac{\sigma}{\sum_{i=2}^{N}(\bar{C}_i^{\beta}-\bar{C}_i^{\alpha})^2/M_i}$)[48], to decrease the coarsening rate, three factors are of mechanistic significance: (1) low diffusivity ($M_i$); (2) reduced interfacial energy ($\sigma$, mediated by coherency); and (3) enhanced partitioning of slow-diffusing alloying elements between the matrix and the precipitates $(\bar{C}_i^{\beta} - \bar{C}_i^{\alpha})^2$. Reducing the interphase misfit strain ($\varepsilon = 100\left|1-\frac{a}{a_0}\right|$) (Eq. 3) is crucial for minimizing the interfacial energy of semi-/coherent $L1_2$ phases, where $a$ and $a_0$ are the lattice parameters of the phase and Al matrix, respectively. A previous study of $L1_2$ strengthened Al alloys has shown that long-term creep strength controlled by interfacial climb is defined with a threshold stress of 40% of Orowan strength (Eq. 1) for particles with an optimum size of 17 nm diameter[49]. Therefore, it is important to achieve an initial dispersion below this optimum size. We thus defined all materials descriptors and their targets to achieve high temperature strengthening (**Fig. 2a, Supplementary Fig. S1**). Next, we define these descriptors to achieve printability.

The LPBF process can be considered as a multi-layer micro-welding process. The main mode of fracture during solidification of Al alloys is hot tearing/cracking, which limits the printability and weldability of many Al alloys such as Al 7000 and 6000 series[50]. Our strategy to avoid hot cracking is to constrain solidification characteristics. This includes limiting the freezing range (FR) and the cracking susceptibility coefficient (CSC) of Clyne and Davies defined by solidification time ratio[51] (**Supplementary Figs. S11a** & **b**). The CSC is best limited by introducing a small fraction of eutectic solidification. The overall hot cracking susceptibility (HCS) was taken as the product of FR and CSC[15], both obtained from Scheil calculations[1] (**Supplementary Fig. S11c**). In total, at. % and size of all phases and coarsening metric of $L1_2$ phases and three solidification parameters: FR, CSC, and HCS are our material descriptors.

**Step II: High-throughput simulations and simple data analytics**

To estimate all above microstructural and solidification material descriptors (**Figs. 2a, 4** & **5**, and **Supplementary Fig. S1**), we generated two sets of 250,000 and 500,000 random compositions using the Latin hypercube sampling (LHS) method (**Supplementary Fig. S4**), with constraints on at. % of {Er, Zr} = [$5\times10^{-3}$, 2], {Y, Yb} = [$5\times10^{-3}$, 1], and {Ni} = [0, 4]. The range has been defined aiming for dilute concentrations of elements and previous experiences[1]. These elements are the inputs for all estimations. It is noteworthy that the model system was made by selecting elements from the periodic table based on their potential ability to form $L1_2$ and metastable phases, their solubility and mobility in Al matrix, their cost, and having control properties such as phase fractions and estimated properties to which we can compare our design and evaluate our design enhancements with our hybrid techniques[1]. Our model material serves as a desired system as the high rate of solidification in AM is beneficial to its properties. Two primary microstructural features that need to be minimized are the inverse of diffusion resistivity (the denominator of Eq. 2) and misfit strain of the $L1_2$ phases (the product is considered as coarsening metric $\frac{\text{Misfit strain}}{\sum_{i=2}^{N}(\bar{C}_i^{\beta}-\bar{C}_i^{\alpha})^2/M_i}$). **Supplementary Fig. S4** shows the diffusion resistivity, misfit strain, and coarsening metric at 250ºC, all normalized by the associated values from the benchmark alloy. Among 250,000 sampled compositions, 28.8% have a lower coarsening metric than the benchmark alloy. The lowest predicted coarsening metric among this population is 2.08X that of the benchmark alloy (**Fig. 2b**), and among the population of 500,000, the maximum is 2.50X. See Supplementary Discussion 1 for more elaboration on the figures.

We find that Zr and Er are the most influential elements on the coarsening metric (**Supplementary Figs. S5-7**, Supplementary Discussions 1 & 2). This qualitative analysis was confirmed by the Spearman coefficient, an index indicating linear correlation for each ranked pair of parameters (**Fig. 4**). Zr decreases coarsening metric as potentially it has the highest partitioning and lowest diffusivity in Al matrix. Er seems has the most negative index on this rate. Strength is also correlated to the $L1_2$ at. % (see Eq. 1); Er is the most influential element increasing this percentage based on the Spearman coefficient analysis (**Fig. 4** & **Supplementary Fig. S8**). While cracking susceptibility is not strongly correlated in a positive or negative manner with composition, FR is strongly correlated with Er and Yb, where increasing Er and reducing Yb lower the range. The Yb influence can be related to the behavior of critical ternary phases, as lowering Yb similarly promotes formation of the eutectic metastable ternary phase. Y and Ni have the highest influence on ternary phase. While this information shows the first rough influence of elements on different microstructural features, it cannot capture non-linear effects of them. We thus need to use ML algorithms to account for these non-linear effects.

**Step III: Develop ML surrogate models**

We chose coarsening metric as a primary microstructural feature and studied the effects of various elements on this descriptor. To develop a surrogate model, we compared six regression techniques: a neural network (NN), K-nearest neighbor (KNN), random forest (RF), support vector machine (SVR), gradient boost (GB), and linear regression (LR) (**Supplementary Fig. S2**). The input vector for our regressors contains the at. % of five alloying elements. Each of these regressors has several hyperparameters which need to be optimized. We used 5-fold cross validation and did the optimization. The process of optimizing the hyperparameters for each technique is described in the Methods Section. We analyzed the test data root mean square error (RMSE) for different training data percentages. As shown in **Supplementary Fig. S2**, the normalized prediction error for our most efficient algorithm, NN, decreases to 3% as the training set comprises of 40 compositions (0.01 %). NN thus requires the lowest amount of data to achieve the lowest error in prediction. Comparing the high difference of NN errors comparing to linear regression (0.2 vs 20) clearly shows that there are non-linear effects which influence our coarsening metrics. The NN surrogate model which connects mole percent of elements to coarsening metrics reliably will be used as a forward prediction model.

**Step IV: Inverse design for various design scenarios to find the optimal design: strategies and uncertainty quantification**

We explored the design space and inversely design the compositions for various target properties. This is the most critical step in the design as it reveals the influential elements on each property and on the combinations of properties while considering non-linear effects. The results can be compared with those in step II to highlight nonlinearities. As shown in **Fig. 2a**, we have two main objectives: high temperature strengthening and printability. This problem has many solutions. The aim is to find the best solution to this problem which leads to highest enhancements in all target metrics/properties. We used BO with a Gaussian process as the surrogate model to perform inverse design. The sampling was done using LHS at each step. To balance exploration and exploitation, we chose expected improvement (EI) as our acquisition function. The detail of this techniques is provided in Methods Section.

The first strategy to solve this problem is to maximize performance and then check if printability metrics are satisfied as well. Thus, initial tests of the BO schema focused on objective functions related to mechanical performance, i.e., targeting high volume fraction of the $L1_2$ phase, small size of this phase, and low coarsening metric compared to the benchmark alloy. Some of the tested objective functions were maximization of (1 / Coarsening metric) and ($L1_2$ / Coarsening metric), followed by integrating the increase in the difference between $L1_2$ at. % from single equilibrium versus Scheil calculations to reduce the size of the phase, (($L1_2^{Single}$ - $L1_2^{Scheil}$)/ Coarsening metric).

We discovered that e.g., for the case of the first objective function, Y and Yb must be zero. Ni has no influence as for each run, BO proposes a new value for the Ni. The coarsening metric is controlled by Er and Zr. High Zr content was also predicted by simple data analytics (Spearman coefficients, **Fig. 4**). However, high Er content is in contradiction to what was concluded above. As seen in **Fig. 4**, the high Er content is due to the influence of Er on $L1_2$ percentage. If Er is not in the composition, all Zr would precipitate as brittle $Al_3Zr$ phases which are detrimental to the strength and ductility. Thus, Er has non-linear effects on the coarsening metric by generating high volume fractions of $L1_2$ phases that then Zr could be dissolved into it and decrease the coarsening metrics. Zr contributes to lower coarsening metrics due to its high partitioning and relatively low diffusivity in Al, and also low misfit strain (see Eq. 2). The objective functions which contain $L1_2$ at. % guides us to the high % of Er to achieve the goal confirming Spearman coefficients. Meantime, the one containing difference of percentages of single equilibrium with the Scheil suggests that again Er is the main influential elements on this objective. Thus, we realized that objective function (1 / Coarsening metric), while we consider Er as a constraint to minimize its percentage in Scheil calculations, is the best objective function to explore. In fact, the exploration resulted in the ML-optimized composition with 40% enhancement in $L1_2$ % versus the benchmark alloy, as well as a 3.5X reduction in the coarsening metric. The ML-optimized alloy met all requirements for printability and all $L1_2$ phases were initially precipitate as ternary phases. Therefore, the ML-optimized alloy was a valid design from all aspects. However, we continued our search with the hope to find compositions with higher enhancements in all descriptors and properties.

The second strategy is to consider printability metrics as the main design objectives, that means minimization CSC, FR, and HCS (**Supplementary Fig. S9**). When evaluating the suggested designs of these iterations, it was seen that the suggested designs were expected to have poor mechanical performance, with higher priority given to features that reduce the metrics with the help of eutectic phases as opposed to strong stable alloys. $Al_3Ni$ typically develops in a eutectic manner near the end of solidification, which enhances printability, but the phase is soft and rapidly coarsens thus reduce the mechanical performance at high temperatures in applications.

For the third strategy of design, we combined a suite of "enhanced" printability metrics with the high mechanical performance targets. Some such objectives were minimization of (CSC* = CSC × Coarsening metric) and (FR* = FR × Coarsening metric) and maximization of ((1 - CSC) / Coarsening metric). The suggested designs for multiple iterations of BO for objectives "CSC*" and "FR*" are presented in **Supplementary Fig. S9**.

Exploring the properties of the designs from these explorations and comparing them to the ML-optimized alloy, it can be observed that, while there are some designs that initially seem viable (with low coarsening metrics and high $L1_2$ fractions), $L1_2$ phases develop at inopportune times, that means having high fractions forming during Scheil solidification. Additionally, these alloys have high fractions of adverse phases, $Al_3Ni$, $Al_3Y$ and $Al_3Zr$, which are connected to poor mechanical performance. $Al_3Y$ can form precipitate in a brittle form ($D0_{19}$ structure) and also at high volume $L1_2$ after aging when triggered by the metastable phase. However, the later results in an increase in coarsening metric and mainly misfit strain due to the high mismatch of Y with Al. We concluded that we could discover guidelines to improve printability; however, when printability is considered in the objective functions, the mechanical performance will significantly sacrifice, and the design fails.

Based on these analyses, we adapted the first strategy and considered minimizing the coarsening metric as our primary goal and put constraint on Er and Zr at. % to control the size and % of $L1_2$ and $Al_3Zr$ phases (**Fig. 2a** & **Supplementary Fig. S1**). For these calculations, for example to obtain the minimum coarsening metric, the BO algorithm was connected directly to ICME techniques, the optimal composition emerged after 4 rounds of sampling with 20 samples in a single batch (80 data). To verify the efficiency of the BO to find the global minimum for this metric, we compared the efficiency with particle swarm optimization (PSO), which is a constrained optimization technique, as an alternative approach and we generated the data using the NN surrogate model. The minimum coarsening metric is again identical to the one from BO for the composition with XNi2Er2Zr at. %. X for Ni indicates Ni does not influence on the coarsening metric. The search converged after eight steps of 10 sampling data (80 sample data). We also rerun BO but this time we did not obtain our data from direct ICME techniques (i.e., calculating coarsening metric after running Thermo-Calc software through TC-python script), and instead we obtained the data through the NN surrogate model. It is noteworthy that if we get data through the NN surrogate model, only 40 data (composition as an input and coarsening metric as a label) to train NN model is the main computational cost (**Fig. 2b**). For BO, the algorithm converged after 100 samples. Interestingly, this shows that the initial generation of the surrogate model, which required 40 samples is helpful in efficiency comparing to 80 sampling data which is required without the surrogate model. In this study, we generated data from simulations which is relatively less costly than experimental data. The comparison shows the efficiency of alternative algorithms in finding extremums and this also depends on the objective functions. The normalized coarsening metric of this optimal composition is 3.5X lower than the benchmark alloy, which shows the power of inverse design techniques compared to high-throughput calculations with random sampling (2.5X after 0.5M data generation) (**Fig. 2b**). Further details regarding the inverse design techniques appear in Methods Section.

As now the coarsening metric is mainly governed by Er and Zr, the problem is reduced to two-dimensional space. The contour plot of the normalized coarsening metric with respect to the Er and Zr (**Fig. 5a**) shows that both are required for improvement in the metric[52]. We also present the contour plots of other microstructural features with respect to these two elements. The contour plots of $Al_3Zr$ at. % and $Al_3Zr$ solvus temperature, which limit maximum Zr content, appear in **Figs. 5d** and **e**. The maximum at. % of Er limits the $L1_2$ at. % and $L1_2$ at. % during solidification and thus increases the volume percentage while the radii remain low through solid-state precipitation (**Figs. 5b** & **c**). We chose the maximum Er at. % equal to 0.4 to limit the at. % during solidification to 1 (**Fig. 5c**). We also considered Zr at. % as 1. We then quantified the uncertainty of both $L1_2$ at. % and coarsening metric and we showed that even with the change of elemental percentages up to 50%, these two descriptors will not change more than 5% with 99% confidence (**Supplementary Fig. S10**). The details related to these analyses is provided in Supplementary Discussion 3.

As discussed, the above single equilibrium calculations provide microstructural features after aging. However, consistent with our experimental observations, non-equilibrium Scheil solidification simulations with our database revealed that the stabilizing $L1_2$ precipitates can appear as a nanoscale metastable ternary phase $Al_{23}Ni_6M_4$ during solidification. The ternary phase acts as a reservoir for the solid-state precipitation of nanoscale $L1_2$ during the aging process. Therefore, we needed to optimize the Ni content to have higher percentages of the ternary phase for a given Er and Zr. For Zr content = 1 at. %, we plotted all the phases previously mentioned from Scheil calculations with respect to Ni at. %. **Figure 5f** shows that initially at Ni = 0 at. %, the ternary and $L1_2$ at. % are zero and a maximum of 1.63 at. %, respectively. As Ni increases, the ternary phase increases to a maximum value of 3.21 at. % at 0.64 at. % of Ni. Thus, Ni has to be > 0.64 at. % while further addition of Ni increases the amount of unnecessary eutectic $Al_3Ni$.

In the next stage, the above-mentioned printability constraints of our design are examined in detail. With Er = 0.4 at. %, for the variation range of Zr and Ni, we plotted CSC, FR, and HCS product (**Supplementary Fig. S11**). The HCS product was very low (< 0.05) for all compositions (compared to 1.13 for Al 7075), indicating good printability within our composition constraints.

**Step V: Rapid experimental workflow to evaluate performance and printability**

We next developed a rapid experimental workflow to validate our predictions for performance and printability in the context of rapid solidification by skipping the powder processing step (Supplementary Discussion 4). We chose a matrix of 9 compositions surrounding the predicted optimum (indicated by diamond points in **Supplementary Fig. S11**). We first fabricated small-scale samples of the nine selected compositions by induction melting, and then mimicked the heating and solidification conditions of AM on these samples by multi-path surface laser scanning. We also performed laser scanning on Al 7075 showing clear hot cracking in association with its known poor printability (**Fig. 2c**). Our prediction was validated with no hot cracks observed for all 9 samples; the sample with Ni = 1.33 at. % and Zr = 1 at. % is shown in **Fig. 2c** for a single laser path. It is noteworthy we also performed a multiple path laser scanning mimicking printing one layer of the powder, and no cracks were detected (**Supplementary Fig. S18**).

Next, to assess the strengthening behavior, we performed Vickers' hardness measurements on these 9 samples in both as-built condition and after aging for 8 hrs at 400ºC. **Supplementary Fig. S12** shows the contour plot of hardness for these 9 samples at these conditions. Consistent with our predicted optimum composition, the sample with Ni = 1.33 at. % and Zr = 1 at. % showed the highest increase in hardness with aging, as expected from $L1_2$ phase precipitation replacing the metastable ternary phase. The aged hardness of 147 HV is 50% higher than that of benchmark alloy, which was processed and tested under identical conditions. Moreover, this hardness is 34% higher than with Zr = 1 at. % and Ni = 0 at. %, which does not contain ternary phases due to the lack of Ni. The SEM-EDS image of this sample shows micro scale Zr rich areas inside grains which are much less present in the sample with Ni = 1.33 at. % (**Supplementary Fig. S16**).

**Step VI: 3D-printing of custom powder and material characterizations**

Based on this validation, we chose Al-2.73Ni-3.19Zr-2.33Er wt% for ultrasonic atomization of the powder (**Fig. 2c** & **Supplementary Fig. S12**). It is noteworthy that based on our uncertainty quantification analyses and as seen in **Fig. 5a**, the coarsening metric of this composition is relatively stable even if the Er or Zr contents changes 50%. The 3D printing process parameters, namely the power and scan speed, were optimized to achieve the highest density as discussed in Supplementary Discussion 5. Samples (28 or $6 \times 6 \times 6$ $mm^3$, **Fig. 3a**) were verified to be crack-free. The EBSD images from the side and top of the sample show non-textured equiaxed grains without employing the challenging addition of nanoparticles in our powder[50] (**Fig. 3b**). Vickers hardness tests parallel to the built direction and on top of the samples were conducted in the as-built condition and after isothermal aging at 400°C. These values are compared in **Supplementary Fig. S17** with the benchmark Al alloy and Al 7075 aged under identical conditions. In as-built condition, the sample exhibits 180 HV, comparable to wrought Al 7075. At peak hardness after 8-10 hrs aging, our alloy has higher strength than Al 7075 and ~50% higher hardness than the benchmark alloy. The hardness has similar gradual overaging at 400°C to the benchmark alloy, while the hardness of the benchmark alloy remains significantly below the designed alloy, confirming its high thermal stability. The hardness also surpasses values from wrought Al 7075 at all times, which is known to be incompatible with LPBF.

We induction melted the design composition. The SEM-EDS image of this sample shows large micro size ternary phases (**Supplementary Fig. S15**). The hardness of this sample is 5 times smaller than the aged 3D-pritned sample, which confirms our concept of design to exploit nanoscale ternary phases for the enhancement of the strength (**Fig. 1**). As mentioned, besides precipitation strengthening, the main strategy which is used in Al alloys is exploiting eutectic phases (dendrite arms) to partition grains[53]. This strategy has significant contribution for the strengthening especially as AM processing decreases the dendrite arms to nanoscale[54]. As seen in **Fig. 3c**, our designed alloy which only exploits precipitation hardening has higher as-built strength to top two designs (benchmark alloy & Al-Ce-Mn) which gets benefit from both mechanisms[1,2]. To further confirm our enhanced strength, we performed tensile tests at room temperatures on samples aged at different aging hours at 400°C. As shown in **Fig. 3c**, the high strength and its stability are also confirmed in these experiments. Once more, our sample has 50% higher yield strength than that of Gong et al.'s benchmark printable alloy[1] ('alloy 1', Al-2.14Ni-1.15Zr-1.35Er-0.25Y-0.73Yb, wt.%).

Microstructural characterization was performed by SEM, STEM, and LEAP APT on the printed material aged for 8 hrs at 400°C, representing the peak hardness condition. The EDS map (**Fig. 6j**) corresponds to the SEM image (**Fig. 6a**) and indicates large scale Ni- and Zr-rich areas consistent with predicted primary $Al_3Ni$ and $Al_3Zr$ phases, respectively. The STEM-HAADF image of precipitates located at the grain boundary (**Figs. 6b** & **c**) shows two distinct structures: the predicted Al-Ni-M ternary phase and Al-Zr/Er $L1_2$ phase (**Figs. 6c** & **d**). Here, the region corresponding to the ternary phase contains Ni- and Er-rich atomic planes, while the $L1_2$ region is primarily Er- and Zr-rich (**Figs. 6d** & **e**). Atomic-resolution STEM-HAADF images of the grain interior show 1-5 nm regions with ordered bright and dim {110} planes, corresponding with $L1_2$ coherent precipitates (**Figs. 6f** & **h**). The Fourier transform of the image shows superlattice reflections consistent with an $L1_2$ structure, and atomic-resolution EDS maps indicate the segregation of Al versus Er/Zr in the $L1_2$ phase (**Figs. 6g-i**). APT results taken from the grain interior clearly confirms the $L1_2$ $Al_3$(Er,Zr) precipitation, corresponding with the ~1-5 nm $L1_2$ precipitates seen from STEM characterization (**Figs. 6k** & **m**). This size scale meets our requirement of starting finer than the optimum particle diameter of 17 nm for creep resistance. The APT of a sample containing ternary phase also confirms a composition near the predicted $Al_{23}Ni_6M_4$ ternary compound (**Figs. 6I** & **n**).

Going forward, nanoprecipitation of metastable phases can be employed to enhance the strength of various alloys. The designed metastable phase can potentially strengthen Ni-based superalloys beside Al alloys. Moreover, this strengthening mechanism can be used for samples manufactured with various rapid solidification processing. The developed hybrid computational workflows can be extended and applied to various multi-objective dilute and multi-component complex concentrated alloy design problems, incorporating other microstructural features contributing to properties and printability, including manufacturing constraints, economic and environmental implications of the alloy composition, and employing accelerated experimental workflows[55,56]. This is especially valuable to leverage the inherent non-equilibrium conditions imposed by rapid solidification in AM, and the opportunity provided by AM for spatially tailored microstructure and multi-material component design in concert with structural optimization methods[57–59]. The algorithms can be enhanced to integrate uncertainty and variabilities in experiment to have robust predictions and to augment different data sources from experiments, simulations, and literature. The interactive framework will work with the alloy designer scientist via multi-agent models[60]. The ful automation of the experimental validation in the future will help to significantly accelerate the structural alloy design cycle of AM and traditionally manufactured alloys for various industrial applications.

## Methods

### Precipitation strengthening

In the Orowan strengthening mechanism, the shear stress required for looping around the precipitate is inversely proportional to the distance between precipitates $\lambda$ (see Eq. 1). In this equation, $M$ is the Taylor factor; $G$ and $\upsilon$ are the shear modulus and Poisson ratio, respectively; $b$ is the magnitude of Burgers' vector; and $\bar{R}$ is the mean planar precipitate radius. Both $\bar{R} = \sqrt{\frac{2}{3}} < R >$ and $\bar{\lambda} = \left(\sqrt{\frac{\pi}{4f}} - 1\right)\bar{R}$ depend on the mean radius $< R >$. Here, $f$ is the volume fraction of the precipitates. Therefore, under a certain precipitate volume fraction, the Orowan strength decreases significantly as the precipitate size increases.

$$\sigma_{\mathrm{or}} = M \frac{0.4Gb}{\pi\sqrt{1-\upsilon}} \cdot \frac{\ln\left(\frac{2\bar{R}}{b}\right)}{\bar{\lambda}} \qquad (1)$$

### Microstructure stability at high temperatures (precipitate coarsening rates)

The sample with minimized coarsening of precipitates maintain the strengthening at higher temperatures[48]. The kinetics of this process is controlled by the volume diffusion of solute elements in small precipitates to adjacent larger ones. Therefore, the coarsening rate depends first on whether the alloying elements in the matrix and precipitates are contributing to bulk diffusion, and second, on how the interface between precipitate and matrix allows this growth of precipitates. The interfacial contribution depends on the interfacial energy. Equation 2 explains the contribution of the bulk diffusion resistivity (denominator) and interfacial energy (numerator) to the coarsening rate in multi-component alloying systems[48]. In this equation, the mobility matrix (***M***) is assumed to be diagonal. The$1/M_i$ are thus the inverse of diagonal coefficients. $\sigma$ is the interfacial energy of precipitate-matrix boundaries, and $V_m^\beta$ is the partial at. volume. $\bar{C}_i^\beta - \bar{C}_i^\alpha$ refers to the temporal average solubility difference of the elements in the precipitates and the matrix at equilibrium and target temperature.

$$k \propto \frac{V_m^\beta \sigma}{\sum_{i=2}^{N} (\bar{C}_i^\beta - \bar{C}_i^\alpha)^2 / M_i} \qquad 2$$

In this manuscript we refer to $\sum_{i=2}^{N} (\bar{C}_i^\beta - \bar{C}_i^\alpha)^2 / M_i$ as diffusion resistivity. Reducing the mismatch of the lattice parameters between a matrix and precipitates is crucial for minimizing the interfacial energy between the phases in both coherent and semi-coherent boundaries. The absolute lattice parameter mismatch for the precipitate and matrix structures is calculated by Eq. 3:

$$\varepsilon = 100 \left| 1 - \frac{a}{a_0} \right| \qquad 3$$

Here, $a$ is a lattice parameter for the precipitate and $a_0$ is the Al lattice parameter. In fact, the lattice parameters depend on both the alloying elements and the thermal expansion resulting from increasing at the target service temperature. We consider the product of misfit strain (ε) and the inverse of diffusion resistivity as our coarsening metric.

**Printability**

The FR is the interval between the temperature at which the alloy is completely liquid and the temperature at which it is 99 % solidified ($FR = T_0^S - T_{0.99}^S$). As this temperature interval get larger, the solidification dendrites are not fed with liquid and leads to shrinkage and thermal contractions and finally cavities and hot cracking. Therefore, minimizing FR is essential to evaluate the HCS. Here, we also analyze HCS of the developed Al alloy by the model developed by Clyne and Davis to evaluate the CSC[51]. According to this model, mass and liquid feeding start from a normalized time $t_{\mathrm{R}}$ (0.6 to 0.1 volume fraction of liquid), in which stresses at the mushy zone can be relaxed. However, the cracking-vulnerable time $t_{\mathrm{v}}$ is defined from 0.1 to 0.01. Consequently, the ratio $t_{\mathrm{v}}/t_{\mathrm{R}}$ results in the CSC parameter. We obtained data for the FR and CSC parameters by conducting nonequilibrium Scheil solidification simulations. To measure the CSC parameter, we assumed that heat flow is proportional to the square root of time. Note that the high-melting precipitation of rare earth elements significantly increases the FR. However, based on experiments, these initial precipitations do not increase HCS. Thus, these precipitates are ignored during the FR and CSC calculations. Therefore, to ensure printability of our Al alloy, we must minimize the HCS, which is the product of FR and CSC[1].

**ML analysis**

Using the LHS random sampling technique, we generated 250,000 and 500,000 data. We then used Thermo-Calc to perform single equilibrium calculations at 250°C[61]. The at. % of the elements in $L1_2$ phases and Al matrix were extracted using the TC-python interface and fed into Eq. 2. The at. % of the five elements in the composition constitutes our feature/input vector. We normalized the input values by deducing the mean and scaling to the variance. We used the Cuml library from RAPIDS[62] developed by NVIDIA for the KNN, RF, SVR, GB, and LR techniques, and TensorFlow for the NN regression technique. We divided the data into training (varying %) and test datasets 20%. The hyperparameters associated with the ML techniques are presented in Supplementary Table S1. They were optimized using grid search and 5-fold cross validation (80% for training and 20% for validation) of the RMSE using gradient descent and Adam optimizers[63]. The target predicted value was the coarsening metric of $L1_2$ phases. After defining the optimum set of hyperparameters for each technique, we measured the prediction error on the test data set for different percentages of the training data. NN regression showed the minimum RMSE compared to other techniques at different percentages. We repeated the measurement 5 times for various initial seeds for NN and the minimum and maximum of the RMSE is shown in **Supplementary Fig. S2** with error bars. The lower RMSE of NN is partially due to the fact that the available hyperparameters in Cuml 3.2 are currently limited and the whole spectrum of hyperparameters is not fully implemented. The NN regressor was thus used to forward predict coarsening metrics used in inverse design techniques. We applied two inverse design approaches.

First, we used BO with GP as a surrogate model and generated data using the NN regressor. As a second method in this approach, we also calculated the labels of the data directly using the TC-python interface of Thermo-Calc. In the second approach, we used PSO for the inverse design technique and generated the labels of data from the NN regressor. GP was chosen as the surrogate model for the BO as it has relatively low number of hyperparameters, our objective function is continuous, and the uncertainty of the fitted model is known. The uncertainty defines the tradeoff between exploration and exploitation in acquisition function of BO.

**SEM characterization**

SEM-BSE and SEM-EDS data were collected on a JEOL JSM7900F FE-SEM equipped with an Oxford Ultim Max 100 $mm^2$ detector. BSE images were collected at 5 kV and 20 kV to optimize surface sensitivity and contrast, respectively. SEM-EDS data were also collected at 5 kV and 15-20 kV, in order to prioritize spatial resolution of surface features at the lower voltage, and to permit detection of the full suite of elements including Er at the higher voltage. EDS spectra and maps were collected and processed using Aztec 5.1 software.

SEM-BSE and SEM-EDS data of zero-nickel and induction-melted alloys were collected on a ThermoFisher Apreo HiVac SEM equipped with an EDAX Elite 150 SDD EDS detector. The BSE-EDS data were obtained at 15 kV in order to measure the presence of all elements and to give the optimal spatial resolution on EDS maps. The EDS spectra and maps were processed by EDAX APEX software.

SEM-EBSD data of the top and side surface of our designed alloy were collected on a ThermoFisher Apreo HiVac SEM equipped with an EDAX Hikari EBSD detector. The EBSD data were collected at 15 kV, and the step size was set to 0.045 µm to secure the spatial resolution. The raw Kikuchi diffractograms were post-analyzed in OIM and home-developed SphInx softwares[64].

**STEM characterization**

Standard lift-out TEM samples were prepared using a Helios Nanolab 660 focused ion beam microscope (Thermo Fisher Scientific). After selecting a sample region containing a representative collection of features of interest via backscatter electron imaging, we extracted a lamella and attached it to a molybdenum Omniprobe TEM grid using an Omniprobe 400 micromanipulator (Oxford Instruments). Due to the large ion range of Ga ions in aluminum, which can cause significant damage and corresponding artifacts during TEM imaging, we reduced the ion beam energy from 30 kV to 16 kV at a sample thickness of 300 nm and then halved it again at corresponding half thicknesses. Final polishing was performed at 750 V using a Model 1040 NanoMill (Fischione Instruments) until a thickness of approximately 50 nm was reached. To remove the effects of Ga-ion damage, the TEM lamella was further cleaned using a Fischione 1051 TEM Mill. Light Ar-ion milling was performed at 0.3 and 0.1 kV for 3 and 1 minutes, respectively.

STEM images were captured with a probe-corrected Thermo Fisher Scientific Themis Z 60–300 kV probe aberration-corrected TEM/STEM using an accelerating voltage of 300 kV, a beam current of 40 pA, and a probe convergence semi-angle of 18.8 mrad. HAADF images were collected with a collection semi-angle range of 78–200 mrad. Atomic resolution HAADF images were drift-corrected using the Revolving STEM (RevSTEM) method, in which 8-frame image series were acquired with a 90° rotation between each consecutive frame[65]. Position-averaged convergent beam electron diffraction (PACBED) patterns were used to determine the sample thickness, which was approximately 30 nm for the imaged region[66]. EDS was performed using a Thermo Fisher Scientific Super X detector and processed using the Thermo Fisher Scientific Velox software. Low magnification EDS was performed using a beam current of 200 pA and filtered using a 5 px averaging filter. Atomic-resolution EDS maps were acquired using a beam current of 50 pA and filtered using non-local principal component analysis (NLPCA)[67].

**APT of 3D-printed samples**

We prepared specimens for APT following standard lift-out methods, using a FEI Helios 660 dual-beam FIB/SEM. APT was performed with a Cameca LEAP 4000 HR, operated in voltage-pulsing mode with the following experimental conditions: base temperature 40 K, pulse rate 100 kHz, pulse fraction 15%, and detection rate 0.5%.

Each LEAP dataset was reconstructed and analyzed using Cameca IVAS software version 3.6.14. We used SEM images of each LEAP tip to measure the shank half angle and tip radius; these images were incorporated into the shape of the reconstruction. Two of the LEAP tips showed a fine dispersion of Zr-rich precipitates, and one LEAP tip showed a large phase rich in Er and Ni. The Zr-rich precipitates were segmented using isosurfaces set to 3.0% Zr (**Fig. 6k**). The Er- and Ni-rich phase was segmented using an isosurface set to 2.5% Ni, as shown in **Fig. 6l**. The reconstructions shown here illustrate the Zr = 3.0% isosurfaces and collection of Er and Ni in second phases, respectively. A proximity histogram concentration profile was applied to both second phases and for all interfaces set by the isosurfaces (**Figs. 6m** & **n**). A bin size of 0.1nm was used for both histograms.

The composition of the matrix and fine precipitates were measured using two LEAP tips segmented with Zr = 0.3 at. % isosurfaces, and the composition of the large second phase was measured from the third LEAP tip (Supplementary Table S2). These measurements indicate that the fine precipitates are the $L1_2$ phase, and the large second phase is the ternary Al-Ni-Er/Zr phase.

**Supplementary Information:**

**Supplementary Discussion 1: High-throughput study of coarsening rate**

To study the coarsening rates of our precipitates, we first performed high-throughput calculations on diffusion resistivity ($\sum_{i=2}^{N}(\bar{C}_i^{\beta} - \bar{C}_i^{\alpha})^2/M_i$), misfit strain (ε), and the coarsening metric, which is ($\frac{\text{misfit } \varepsilon}{\sum_{i=2}^{N}(\bar{C}_i^{\beta} - \bar{C}_i^{\alpha})^2/M_i}$). **Supplementary Fig. S4** shows the distribution of the input elements for all generated data, which have a good uniform distribution in the studied range. The three target parameters were calculated; the values were normalized with the associated values from the benchmark printable Al alloy (alloy 1)[1], which has the same alloying elements and is designed for the identical combination of properties.

**Supplementary Fig. S5** shows the distribution of diffusion resistivity with respect to the studied parameters: Ni, Er, Zr, Y, and Yb at. percentages; diffusion resistivity; misfit strain; and the coarsening metric. **Supplementary Fig. S5a** shows that lowering the amount of Ni suppresses normalized diffusion resistivity above 1.6 and below 0.1. Moreover, Er decreases the diffusion resistivity when increased to 0.35 at. %. Further, high diffusion resistivities are for Er in 1.2-1.6 at. %. Moreover, it is evident that Zr has a notable contribution to the increase in diffusion resistivity. Analysis of the diffusion resistivity with respect to Y at. % shows that the highest values are achieved when Y at. % ranges between 0.2 and 0.3. When Yb is increased, the diffusion resistivity decreases initially from the maximum values throughout the range and appears to distribute uniformly in the 0-1.65 range. **Supplementary Fig. S6** shows the distribution of misfit strain with respect to the studied parameters. The distributions do not indicate a specific correlation between the at. % of alloying elements and the misfit strain. The distribution of misfit strain with respect to coarsening metric shows that at lower coarsening metrics, the misfit strain also achieves low values, as expected.

**Supplementary Fig. S7** shows the coarsening metrics with respect to the studied parameters: Ni, Er, Zr, Y, and Yb at. %; diffusion resistivity; and misfit strain. These subplots reveal that among alloying elements, Zr plays the most significant role in decreasing the coarsening metric. Moreover, the initial increase of Er concentration from 0 to 0.4 at. % seems to increase the range in the variation of coarsening metrics remarkably. **Supplementary Figs. S7g** and **S7h** show that the coarsening metric is highly influenced by diffusion resistivity and is less influenced by misfit strain.

We also analyzed $L1_2$ at. % with respect to studied parameters. As shown in **Supplementary Fig. S8**, Er has the maximum contribution for the increase of this phase. In contrast, an increase in Ni and Y decreases this phase.

**Supplementary Discussion 2: Spearman coefficients**

The correlations between the studied parameters are shown in the heat map in **Fig. 4**. The heat map represents the Spearman coefficient and indicates a linear correlation for each ranked pair of parameters. The Spearman coefficient values are also presented in this figure. The parameters show that, as expected, Zr has the greatest influence on the diffusion resistivity, increasing it by 0.78; Yb has the inverse effect, decreasing it by 0.35. Er and Y rank second and third, respectively, in terms of their influence on decreasing the bulk diffusion resistivity. Ni seems to have no influence on this parameter. The fluence of these elements on coarsening metric is the same but in reverse trend of their diffusion resistivity index. Finally, as can be seen from the distributions in **Supplementary Figs. S7g** and **S7h**, the diffusion resistivity is more influential, decreasing the coarsening metric (by 0.96) than the misfit strain (by 0.22).

FR is mainly controlled by Yb, Er, and Ni. Yb negatively influences formation of eutectic ternary phases and Ni influences on both eutectic ternary and $Al_3Ni$ phases. Er and Y have the highest effects on CSC, while none of them can strongly change this parameter. To reduce the HCS, additions of Er and Yb has the highest advantageous (by 0.51) and disadvantageous (by 0.50) based on the Spearman coefficients.

**Supplementary Discussion 3: Uncertainty quantification for performance descriptors**

Uncertainty quantification was performed using the t-statistics test, a statistical measure for hypothesis testing to determine whether a process to change has some impact on a feature of interest. A method for this is the "one-tailed test" which determines if, for a distribution, the mean value of a target value is greater than or less than an expected mean. This test is conducted by calculating the critical "t" statistic as $t = \frac{\hat{\theta} - E(\hat{\theta})}{STD(\theta)}$, where STD(θ) is the standard deviation of the distribution, $\hat{\theta}$ is the calculated mean of the sample set, and $E(\hat{\theta})$ is the expected mean value. The tailed test relies on a null ($H_o$) and alternative ($H_A$) hypothesis and compares the calculated statistic against a known value (z), given the confidence interval and dataset size. The relation of the calculated and known value determines if the null hypothesis can be rejected or accepted.

For this work, the test was applied to predict if variations in composition off the ML-optimized design would have significant impacts on performance (as coarsening metric and fraction of $L1_2$). The dataset was constructed by forming three normal distributions for Ni, Er, and Zr, treating the composition of the design as the mean, and a 50% change as the standard deviation. 1250 values were sampled from each distribution to form the dataset, which were then modeled our high throughput framework.

The t-test was performed in two manners: by treating $H_o$ as the mean $L1_2$ fraction being greater than or equal to 95% that of the optimized design, $\mu_{L12} \geq 0.95\ L1_{2,ML}$, (and $H_A$ as being less than this value), and with $H_o$ as the mean coarsening metric being less than or equal to 105% the optimized design, $\mu_{Cmetric} \leq 1.05\ \ Cmetric_{ML}$. This structure was chosen to predict if a 5% loss in performance is expected to occur on average.

After simulating the coarsening metric and $L1_2$ fraction of the 1250 designs, the statistic for each distribution was calculated using the above equation and compared to the confidence interval for this size of a distribution and 99% confidence (+/-2.33). The calculated statistic for $L1_2$ values was found to be greater than this, indicating that with 99% confidence the null hypothesis could not be rejected (it could be accepted that under the 50% variance in composition the mean fraction of $L1_2$ is greater than or equal to 95% of the optimized design). Similarly, the coarsening metric was significantly below the critical value, as such the null hypothesis could not be rejected. The distributions of the simulated $L1_2$ fraction and coarsening metric are presented in **Supplementary Fig. S10**.

**Supplementary Discussion 4: Experiments on laser-scanned induction-melted samples to validate printability and precipitation hardening**

For the experimental investigation of printability, we demonstrated an approach to assessing the development of hot cracks as a function of chemical composition. First, we scaled the 9 batches determined via ML using an XPR Analytical Balance (Mettler Toledo) with an accuracy of 0.005 mg. The purity of the raw elements utilized in this study exhibited was as follows: Al (99.99%), Ni (99.99%), Er (99.9%), and Zr (99.95%). The elements were employed in granulate shape (<5 mm). For each batch, a total of 30 g was cast using the induction heating system MC20V (Indutherm) to ensure a homogeneous elemental distribution in the $Al_2O_3$-crucible.

Before heating, a vacuum was pumped to 0.1 mbar, reducing residual $O_2$, then afterward flooded with Ar to normal pressure. The melting chamber was subsequently flooded with Ar sequentially to avoid oxidation during melting. The overheating temperature of the melts was adjusted to 1000°C, at which full melting of raw elements was achieved. The holding time at this melting temperature was set to 5 min. The castings solidified as rods with a length of approximately 60 mm and a diameter of 15 mm. These rods were rolled to obtain sheets. The roll gap was iteratively reduced over 10 steps to the desired thickness of 2 mm. During the rolling procedure, the specimens were heated to 100°C to avoid edge cracking. Using this procedure, we obtained sheets with a dimension of 280 × 35 × 2 $mm^3$. Because rolling caused the sheets to bend marginally, we performed a pressing step to produce flat sheets. The sheets were then laser scanned using an LT30 LPBF machine (DMG MORI) on multiple paths mimicking one layer printing of the powder. To ensure its performance and reliability, we generally calibrated the LPBF machine according to ISO/ASTM DIS 52941:2019. This system is equipped with a solid-state Nd: YAG laser source with a wavelength of 1064 nm and a maximum laser power of 1000 W. Additionally, the laser beam had a Gaussian distribution with an adjusted laser focus of 70 µm. Regarding the atmosphere in the build chamber, Ar -4.6 was used to reduce a residual $O_2$ content to <1000 ppm.

The baseplate temperature was set to room temperature and the baseplate was sprayed with BN to avoid a potential joining of the sheets to the baseplate. The LPBF processing parameters were used to ensure a deep penetration (~1.2 mm) by the laser. The following combination of process parameters was determined based on preliminary studies: scan speed 300 mm/s, hatch distance 170 µm, and laser power 400 W. A region of 200 × 25 $mm^2$ was laser-scanned employing a stripe-scanning strategy with a vector length of 8 mm to minimize warpage and residual stress in the Al-sheets. Both sides of the sheets were laser scanned to ensure the rapid solidification of the microstructure throughout the processed volume. To validate the development of hot cracks during laser scanning, the specimens were extracted from the laser-exposed region. The cut specimens were conductively embedded, ground, and polished using a Hexamatic (Struers). Next, they were vibro-polished using a Vibromet (Bruker) via colloidal suspension. Light micrographic images were then generated using a VHX 5000 (Keyence) to detect undesired process-induced defects.

The mechanical properties of the laser-scanned sheets in as-built and aged for 8 hrs at 400°C were analyzed. As part of this process, micro-hardness of the sheets was measured using an automatic Vickers hardness tester FA 30 (KB) applying 100 gram force loading. At least five measurements on the same sample were performed and the average value is reported (**Supplementary Fig. S12**). We further performed Vickers hardness tests on induction melted sample of our final design: Ni = 1.33 at. % and Zr = 1 at. %. Ten measurements on the same sample were performed and the average value and the standard deviation are shown in **Supplementary Fig. S17**.

**Supplementary Discussion 5: 3D-printing of custom alloys**

The powder materials were ultrasonically atomized using an AUS500 system (Indutherm Bluepower). For the induction melting, an $Al_2O_3$ crucible was coated with hexagonal boron nitride paste to prevent the melt from reacting with the crucible material. The obtained particle size distribution was measured by laser diffraction method. The particle size distribution for the ML-optimized powder material was as follows: d10 = 44.5 %, d50 = 63.8 %, and d90 = 91.5 %. The chemical composition of the powder material was measured for the ML-optimized Al-2.47Ni-3.25Zr-2.38Er, which was detected via X-ray fluorescence measurement.

Subsequently, specimens were additively manufactured via LPBF using an SLM Solutions 250 HL machine. This machine has an Nd:YAG laser with a maximum laser power of 400 W operating at a wavelength of 1064 nm. We applied a Gaussian beam distribution with a laser beam diameter of 70 μm. The following processing parameters were employed: laser power 350 W, hatch distance 120 μm, scan speed 1100 mm/s, layer thickness 50 μm, scan strategy 8 mm stripes, and rotation of 67°. Regarding the build chamber conditions, a preheating temperature of 200 °C was selected, and during 3D-printing, Ar-4.6 was employed as an inert gas atmosphere with a residual $O_2$-level of < 1000 ppm. Additionally, before SLM processing, the powder materials were vacuum dried to reduce their relative humidity by < 5 % using a system developed in-house. The specimen geometry printed were with the dimensions of 6 $mm^3$ and 2 cm × 6 mm × 6 mm. The chemical composition of the printed ML-optimized specimens was Al-2.55Ni-2.9Zr-2.37Er detected via X-ray fluorescence measurement.

**Supplementary Discussion 6: Microscopy and mechanical performance of 3D-printed samples**

**STEM characterization:**

STEM high-angle annular dark-field (HAADF) imaging and energy dispersive X-ray spectroscopy (EDS) confirms a microstructure consisting of 1-2 µm Al grains with grain boundary precipitates (20-50 nm) rich in Ni and Er scattered throughout the aged sample, along with larger Ni-rich regions corresponding to the $Al_3Ni$ intermetallic (**Supplementary Figs. S13** & **S14**). Atomic resolution imaging of these precipitates reveals a non-uniform composition and structure coherent with the "parent" grain, in this case imaged along a <110>-type zone (**Fig. 6c**). Significant Z-contrast differences indicates two distinct structures whose relative chemical distribution is supported by EDS: type (1) consisting of single Er-rich atomic planes spaced by Ni- and Al-rich regions of approximately in **Fig. 6f**, and type (2) consisting of Er- and Zr-rich planes arranged in an $L1_2$-type ordering with Al (**Figs. 6b-i**). The differences in relative chemical compositions suggest that these structures correspond with the predicted Al-Ni-Er ternary and the Al-(Zr/Er) $L1_2$ phases, respectively, as shown by EDS maps in **Figs. 6d** & **e**. In addition, similar $L1_2$-type structures are found attached to or near the ternary phase for several of these grain boundary precipitates (**Figs. 6c-e**); combined with the expected transition from the ternary to the $L1_2$ phase during aging, this result suggests incomplete transformation.

From HAADF STEM images, the formation of sub-10 nm nanoprecipitates throughout the Al matrix is shown as well, confirming simulation results. Atomic-resolution images of the grain interior show alternating contrast in atomic planes consistent with $L1_2$-type ordering between higher Z and lower Z elements, with the Fourier transform of the image confirming the 001 superlattice reflections (**Figs. 6f** & **g**)[65,66]. In contrast with the grain boundary, the Ni concentration in the matrix is low, suggesting that these nanoprecipitates are primarily Al-Zr/Er. This has been confirmed by STEM EDS (**Figs. 6h** & **i**).

The mechanical properties of as-built and aged 3D-printed samples were analyzed. As part of this process, micro-hardness of the sheets was measured using a Vickers hardness tester applying 100 gram force loading. The reported hardness is along build directions and from the top surface of the samples. The samples were aged at 400°C for various hours and their hardness were measured and compared with wrought Al 7075 aged at identical conditions. For all these experiments, the average for each aging hour and the raw data are shown in **Supplementary Fig. S17**.

The as-built and aged tensile samples cut using wire electrical-discharge machining (EDM). The thickness = 1.5 mm, width = 7 mm, length = 2.4 cm, gauge length = 1 cm, and gauge thickness = 3 mm. The samples were loaded at room temperature based on ASTM E8 standard in Westmoreland Company. The samples were deformed with 0.0178 cm/min. Optical extensometers were placed on opposite sides of each specimen to measure axial strain throughout testing.

**Acknowledgements**

This work was supported by the MIT-Portugal Program with funding provided to A.J.H. We thank Stephan Kraemer, Austin Akey, and Timothy Cavanaugh from Center for Nanoscale System at Harvard University who helped us with FIBing samples for STEM and APT, and also SEM characterization. We acknowledge Amazemet Company for atomization of the custom powder. We thank QuesTek Company for providing us the database for CALPHAD simulations. M.X. and J.M.L. acknowledge support from the National Science Foundation (No. CMMI–1922206) for STEM characterization. The STEM characterization utilized the MIT.nano Characterization Facilities. G.B.O. thanks Thermo-Calc for free licensing at MIT in conjunction with his Thermo-Calc professorship. F.H. and M.S. acknowledge the German Research Foundation (Deutsche Forschungsgemeinschaft) supporting the utilized DMG MORI LT30 machine under contract No. INST 214/228-1 FUGG. Z.G. acknowledges the funding from Army Research Laboratory. S.M.T.M. thanks Tonio Buonassisi, Kaihao Zhang, Bo Ni, and Matthew Melfi for fruitful discussions.

**Author contributions**

S.M.T.M. conceived and designed the project. S.M.T.M. designed the hybrid frameworks. S.M.T.M and G.B.O. designed the ICME techniques. S.M.T.M., A.J.H., and G.B.O. designed the experiments. S.M.T.M. developed ICME, ML and hybrid libraries and conducted all simulations and data analytics. S.M.T.M. performed indentation on 3D-printed and wrought Al 7075. M.X. and J.M.L. performed TEM and STEM analysis. F.H. and M.S. performed induction melting and laser scanning experiments and conducted their indentation and optical microscopy analysis. F.H. and M.S. performed LPBF using the custom alloy powder. C.H. performed LEAP APT. G.Z. performed SEM-EDS and EBSD on induction melted and laser-scanned samples with guidance from S.L.W. G.Z. performed sample preparation of the tensile samples. B.G. performed uncertainty quantification analysis and contributed to test different strategies and performing high throughput Scheil simulations. S.M.T.M. and G.B.O. analyzed the results. All authors participated in discussions about the results and contributed to interpretation of the corresponding results. S.M.T.M., G.B.O., and A.J.H. wrote the paper and received input from all authors.

### Competing interests

S.M.T.M., A.J.H., and G.B.O. are co-inventors on a utility patent application (PCT/US2023/078228) with an International Publication Number of WO 2024/092273 A2 that includes the numerical methodology, the alloy compositions, processing conditions, and the microstructural features described in this paper.

### Data availability

The data for performing ThermoCalc analysis on Al alloys is owned by QuesTek Company. All data produced by simulations and experiments and the parameters for the data generation, which are required to reproduce the findings during this study, are included in this manuscript and under Supplementary Information.

### Code availability

The in-house code for hybrid ML/CALPHAD simulations has restrictions for accessibility due to utility patent application. The CALPHAD part of the code was performed using ThemoCalc software, which is not publicly available. However, an essential part of the code which works independently and a tutorial on all major code connections and design optimization has been provided attached to this manuscript and will be provided online when the paper is accepted.

## Main figures:

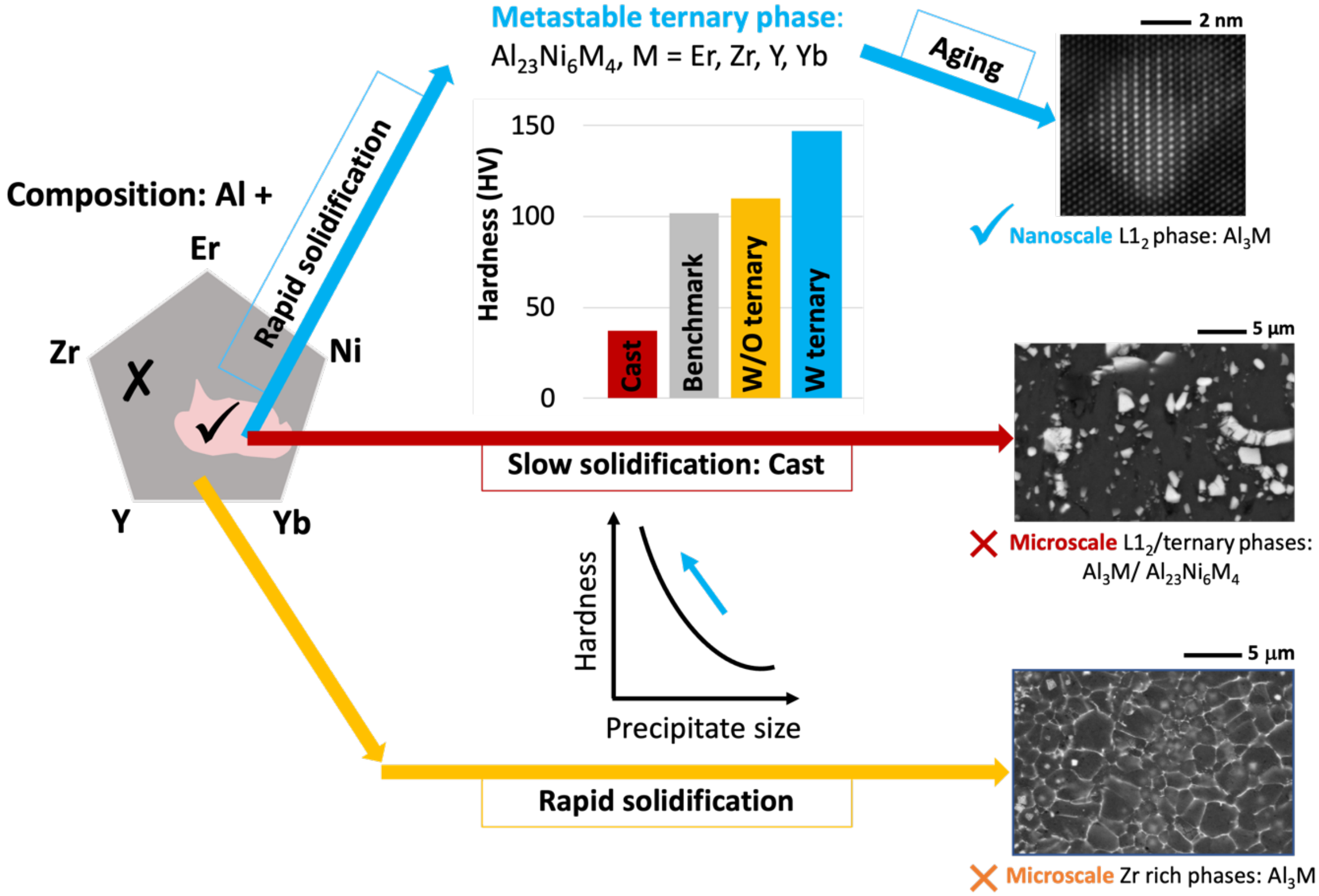

Fig. 1. a) Alloy design concept showing how exploiting nanoscale metastable phases due to rapid solidification and correct choice of composition transforms the length scales of the hardening phase from micro to nanoscale. TEM images showing nanoprecipitation of $L1_2$ phases inside grain interiors (blue pathway). If the same composition undergoes slow solidification such as cast, then microscale strengthening phases appear and the Vickers hardness is five times lower. If rapid solidification is applied on a composition which does not allow precipitation of metastable phase (yellow color), once more the hardness is ~30% lower than the blue design pathway. Microscale Zr rich regions appear in grain interiors.

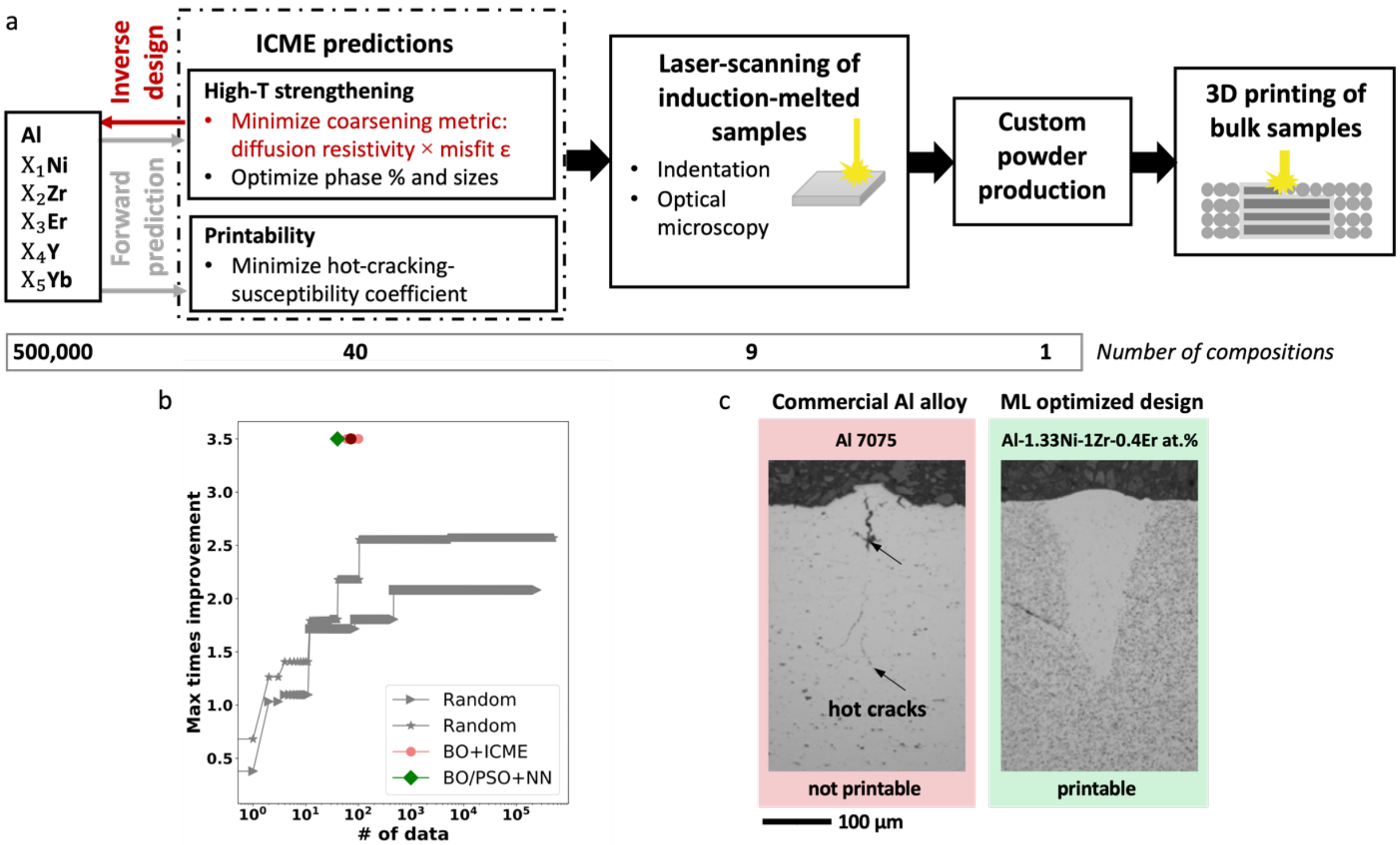

Fig. 2. a) The combined numerical and experimental workflow of our design. We started from 500,000 random data. Using various ML and inverse design algorithms and 40 sampling data, we down-selected 9 compositions. The prediction considered both the performance (high-temperature (T) strengthening) and printability of the design. The printability and hardness of nine laser-scanned induction-melted compositions were analyzed in the next step, and one composition was down-selected. A custom alloy from a selected composition was developed, and the sample was 3D-printed. The hardness of the sample was measured at different aging hours. b) Maximum coarsening resistance achieved by different inverse design techniques and the required amount of data. The average required number for BO+ICME is shown with darker maroon circle and the light circles show the data for five trials. The evolution of maximum coarsening resistance is shown for a random sampling of 500,000 and 240,000 data. c) The optical images of laser-scanned induction-melted samples of Al 7075 and our ML-optimized alloy show that the latter does not suffer from hot-cracking.

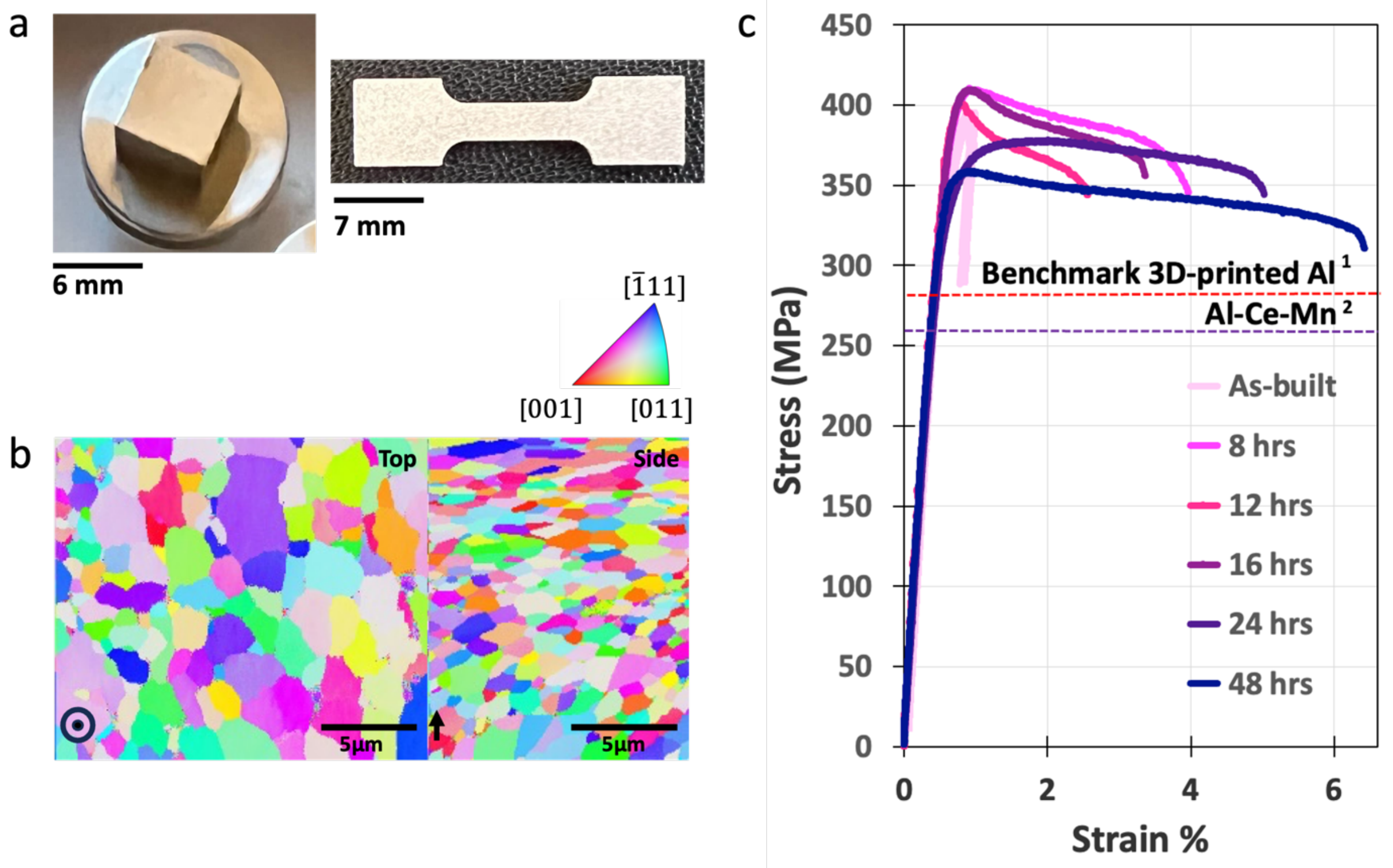

Fig. 3. a) A cube of 3D-printed samples and a tensile sample from the ML-optimized alloy are shown. b) EBSD images from the top and side of the 3D-printed cube. The built directions are shown on the images. c) Room temperature tensile stress-strain curves of as-built and aged samples at different aging hours at 400°C. The yield strength of the benchmark printable Al alloy is shown. The strength of Al-Ce-Mn, which has a highest reported hardness in current AM and cast alloys is also shown.

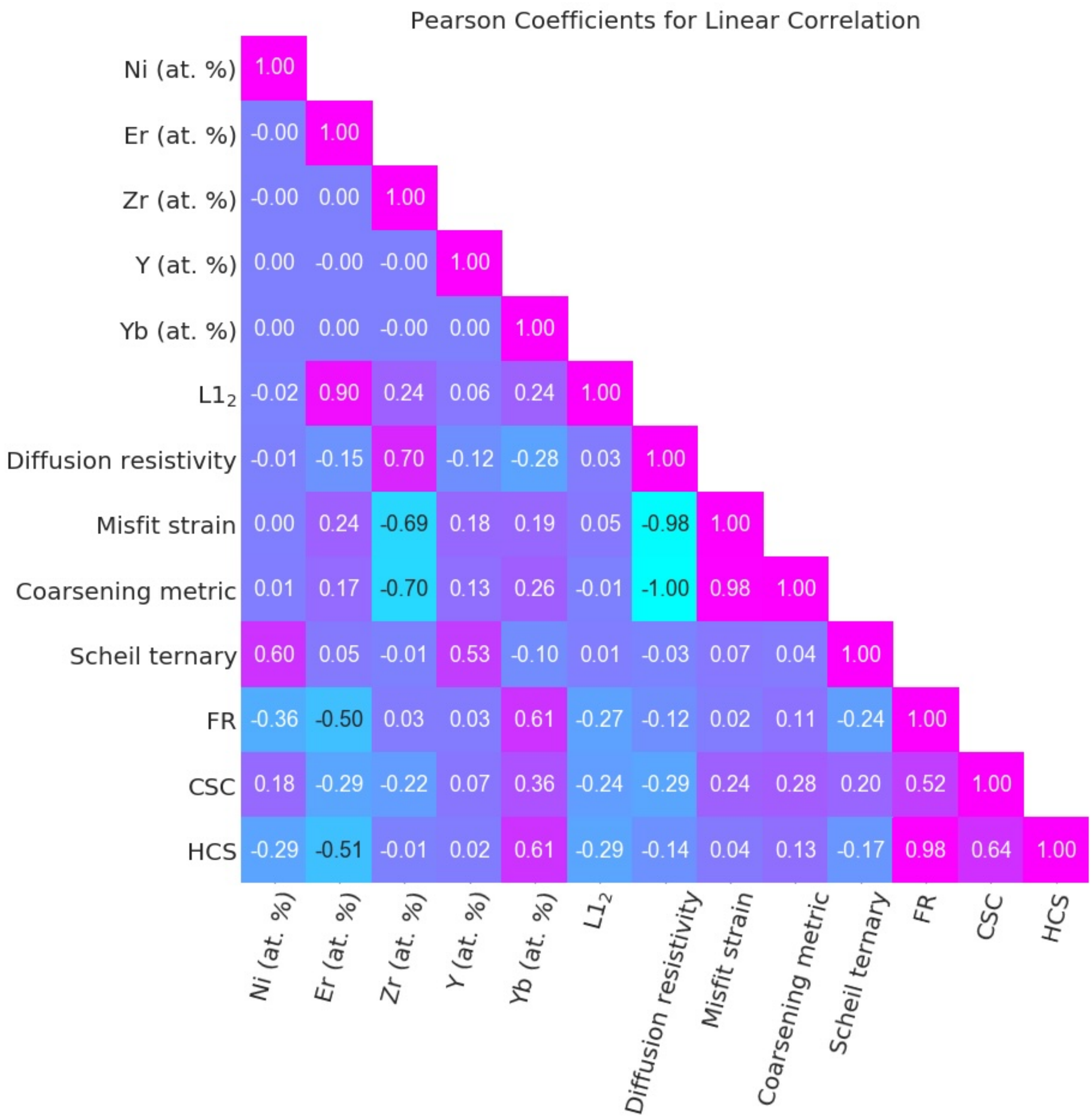

Fig. 4. Heat map indicating the Spearman coefficients between each pair of parameters in the design space of the model alloy Al-Er-Zr-Y-Yb-Ni. The descriptors are calculated from high-throughput single equilibrium or Scheil simulations. Coefficients demonstrate strong dependence of metastable ternary phases on Ni and Y constituents, while the addition of Zr and Er are most strongly correlated with low coarsening metric and high fractions of L1$_2$ phase, respectively. FR and HCS depend strongly on Er and Yb.

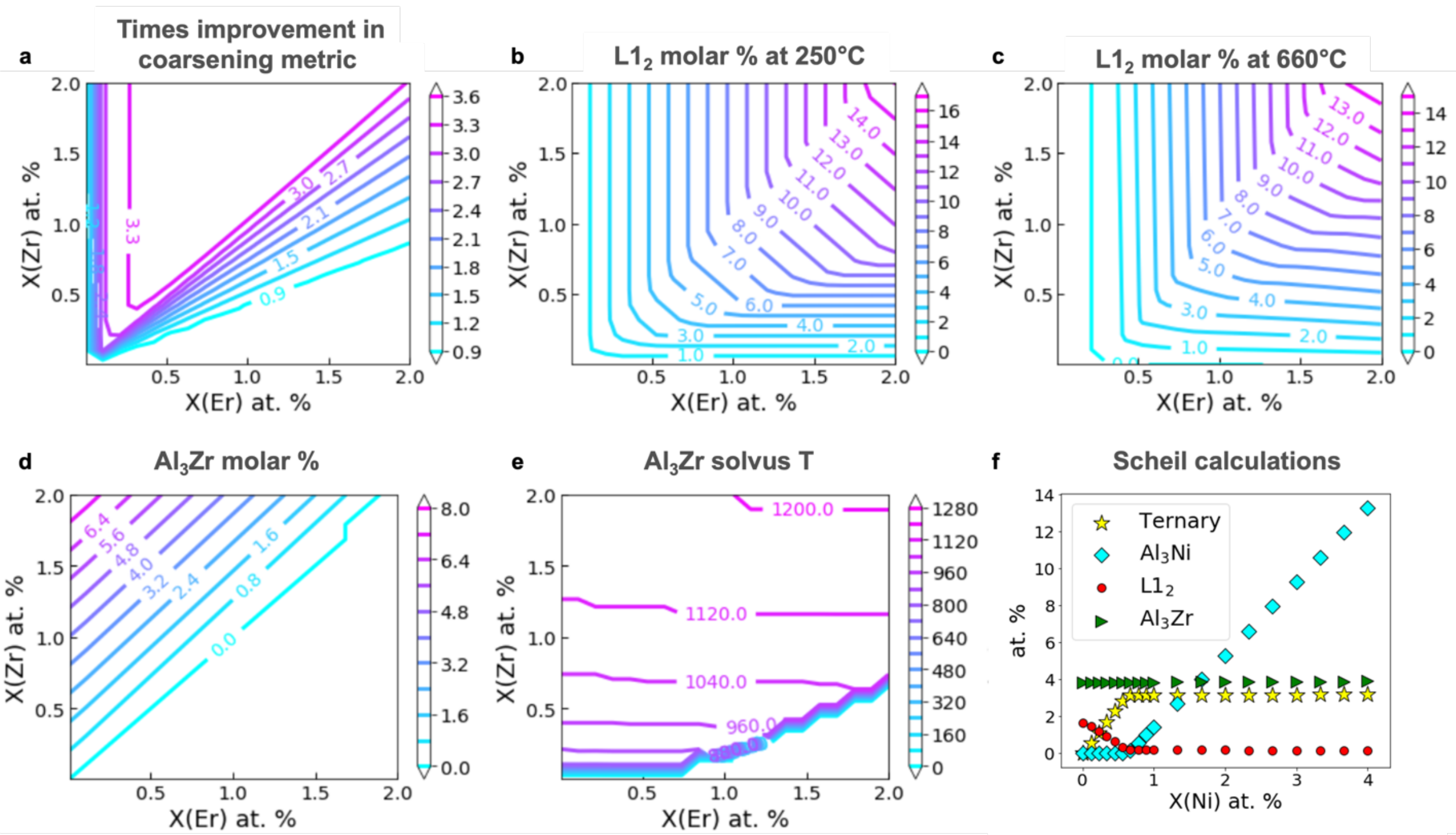

Fig. 5. a-f) Forward prediction of structural parameters with respect to Er and Zr at. %. Each structural feature defines a limit on the Er/Zr content. f) At. % of different phases produced by changing the Ni at. % in Scheil simulations (Er = 0.4 at. % and Zr = 1 at. %). The reduced space of "a-f" only with respect to Er and Zr and "g" only with respect to Ni, are explained in the Results and Discussions Section.

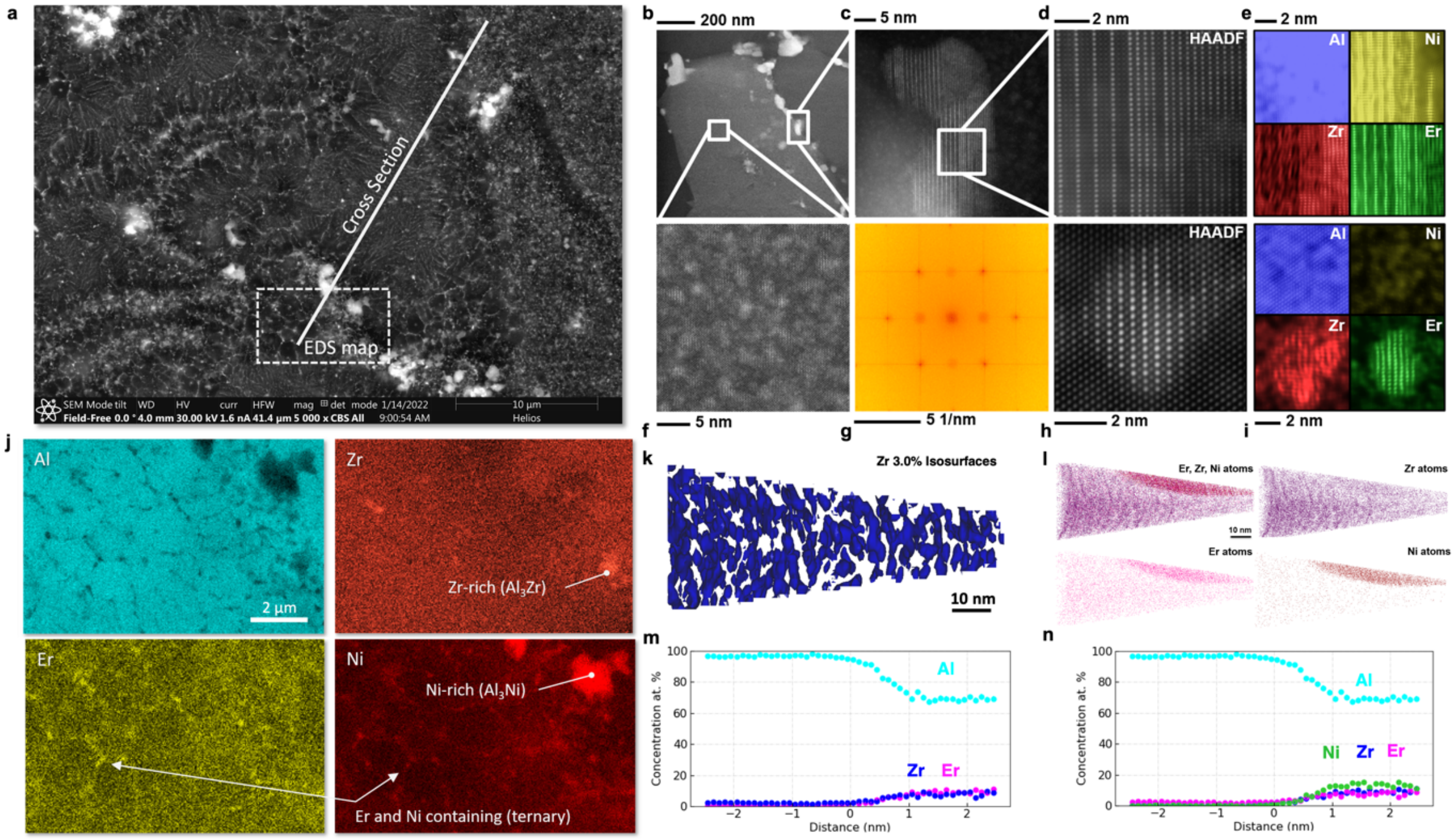

Fig. 6. a) SEM image of our ML-optimized 3D-printed sample (aged at 400°C for 8 hrs) shows various shading (phases) in the microstructure. b) HAADF STEM overview of the imaged grain along a <110>-type zone. c) Higher-magnification image of a grain boundary precipitate coherent with the parent grain; d) atomic-resolution HAADF STEM image of this precipitate; and e) corresponding de-noised EDS Al/Ni/Zr/Er elemental maps. f) HAADF STEM image of the grain interior showing $L1_2$ nanoprecipitates as suggested by g) {001}-type superlattice reflections present and circled in the FFT. h) Atomic-resolution HAADF STEM image and i) corresponding de-noised EDS Al/Ni/Zr/Er elemental maps for an example $L1_2$ nanoprecipitate in the interior of the grains. j) EDS image of SEM image showing elemental distribution. k) Reconstruction of LEAP dataset containing a fine dispersion of Zr-rich precipitates. The isosurfaces where Zr=3.0% are shown in blue. l) Elemental segregation of Er and Ni to large secondary phase. m) Proximity histogram for isosurface interfaces where Zr=3.0%, showing enrichment in both Zr and Er. n) Proximity histogram for isosurface interfaces where Ni=2.5%, showing enrichment in Ni, Er, and Zr.

## SI figures & tables:

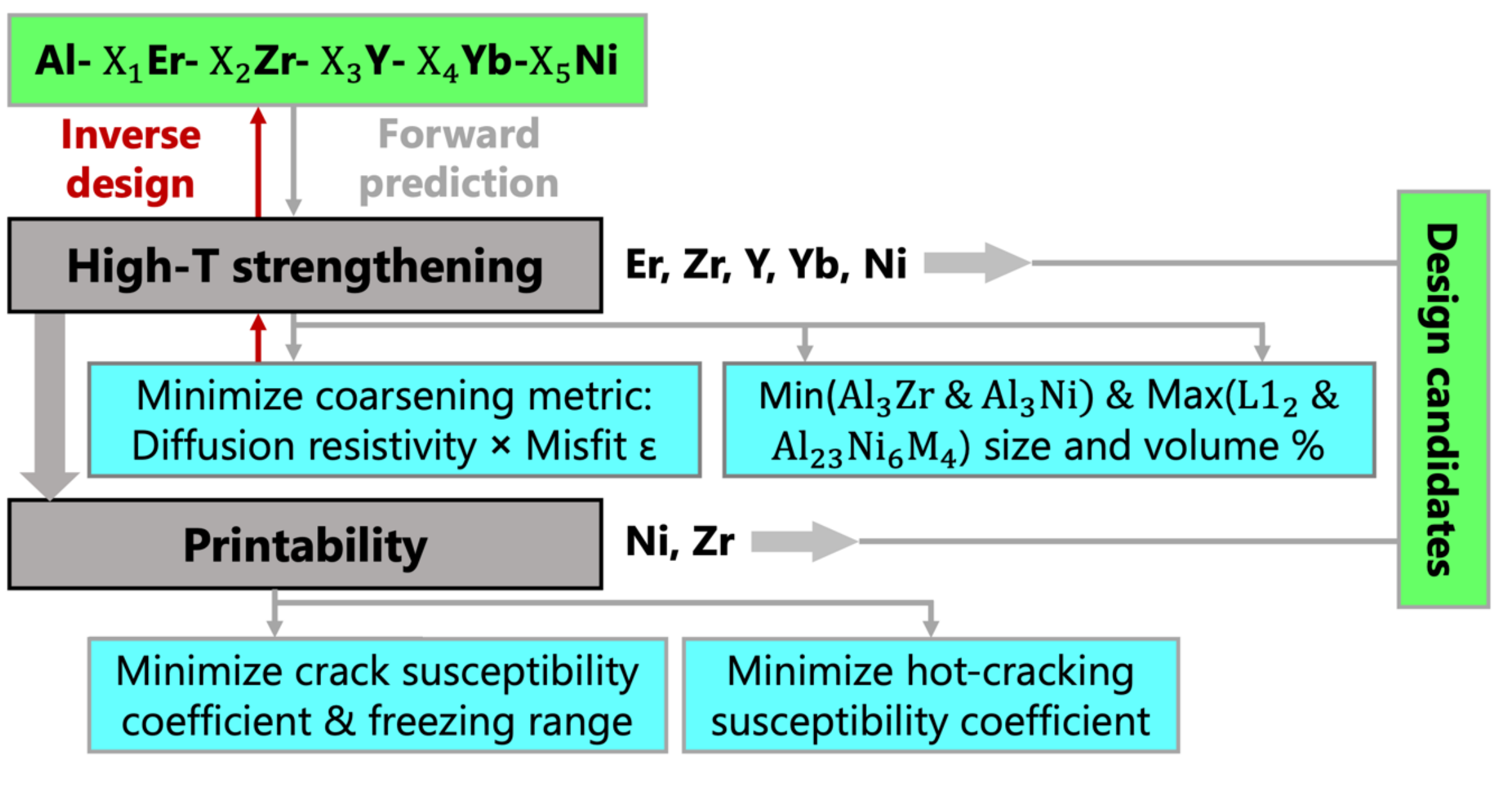

Fig. S1. Flow diagram and design approaches for measuring structure parameters. The composition is inversely designed for the primary goal of minimizing the coarsening metric of $L1_2$ phases. All other microstructural features are forward predicted and work as constraints for the inversely designed composition.

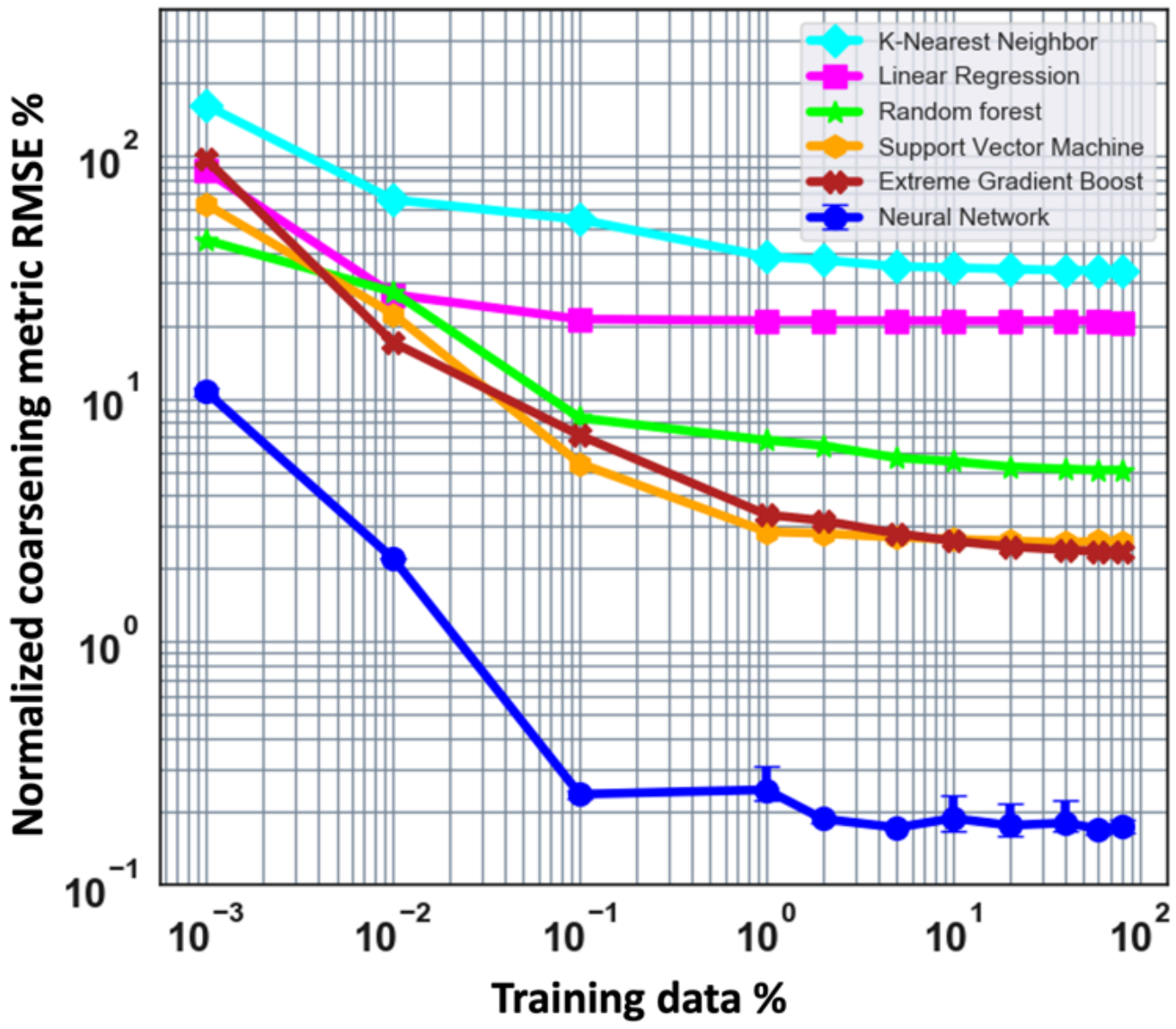

Fig. S2. Root mean square error (RMSE) of the normalized coarsening metric with respect to the training data % for six applied ML techniques: neural network (NN), K-nearest neighbor (KNN), random forest (RF), support vector machine (SVM), gradient boost (GB), and linear regression (LR) techniques. A minimum of 40 samples is required to achieve < 3% of error to predict the entire compositional space for the most efficient algorithm, NN. For NN the simulation was repeated for ten random initial seeds and the standard deviation is shown.

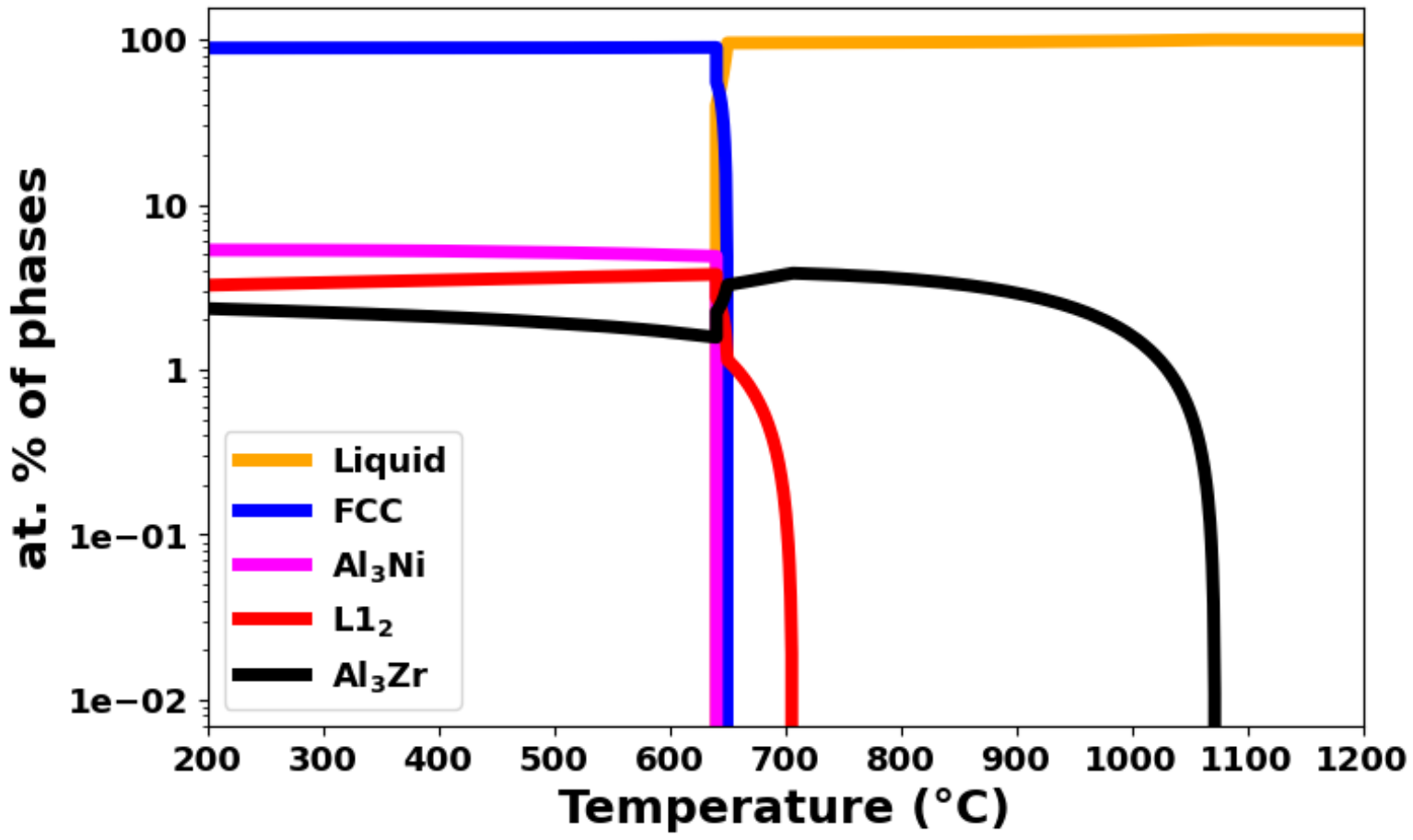

Fig. S3. CALPHAD simulation of phase at. % with respect to temperature for our designed Al-Ni-Er-Zr alloy.

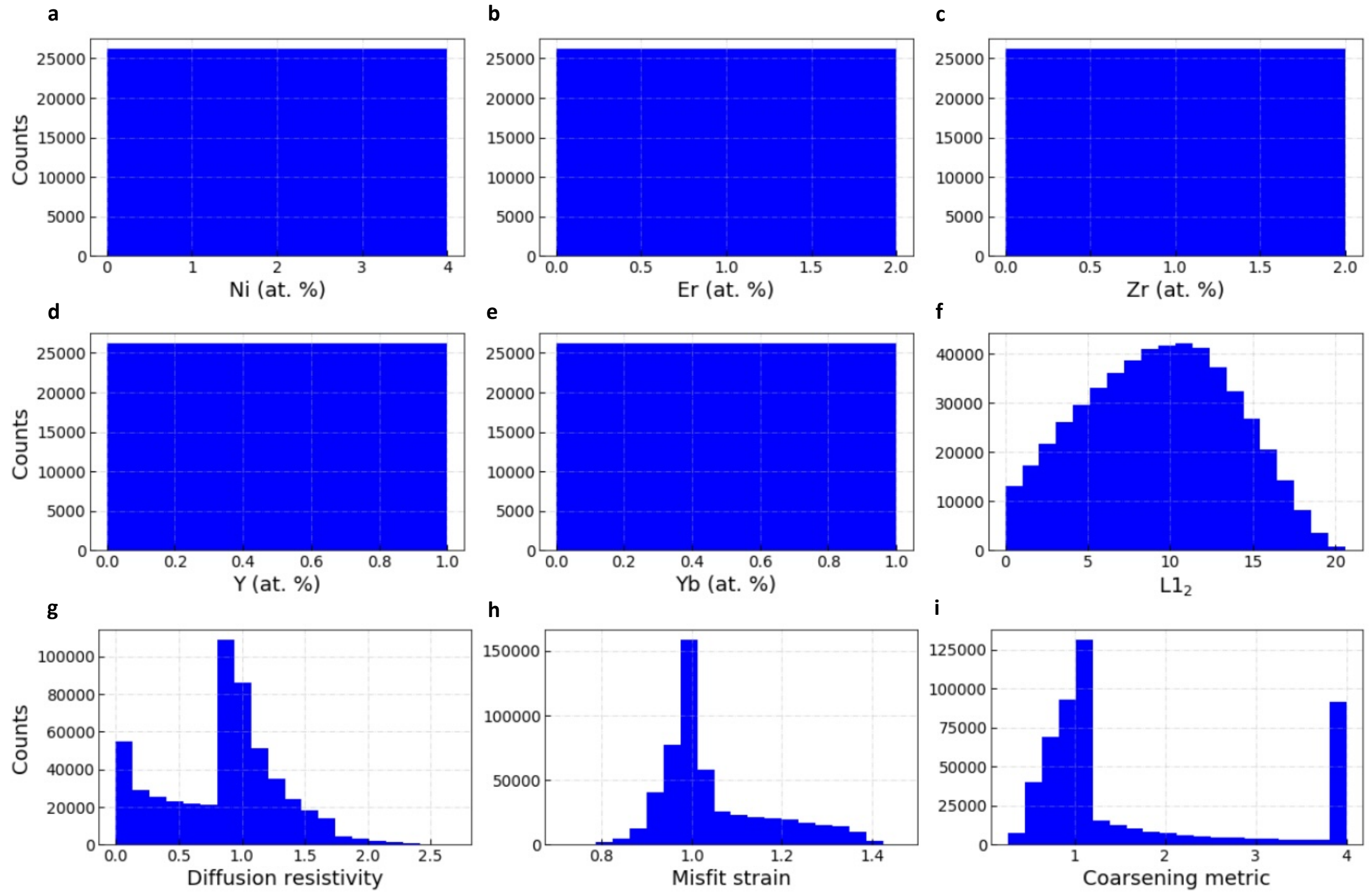

Fig. S4. Distribution of at. % for alloying elements and normalized target properties (i.e., L1$_2$, diffusion resistivity, misfit strain (ε), and coarsening metric) for 500,000 random compositions. Target properties are normalized with associated values from the benchmark printable Al alloy. Coarsening metrics > 4 are assigned 4.

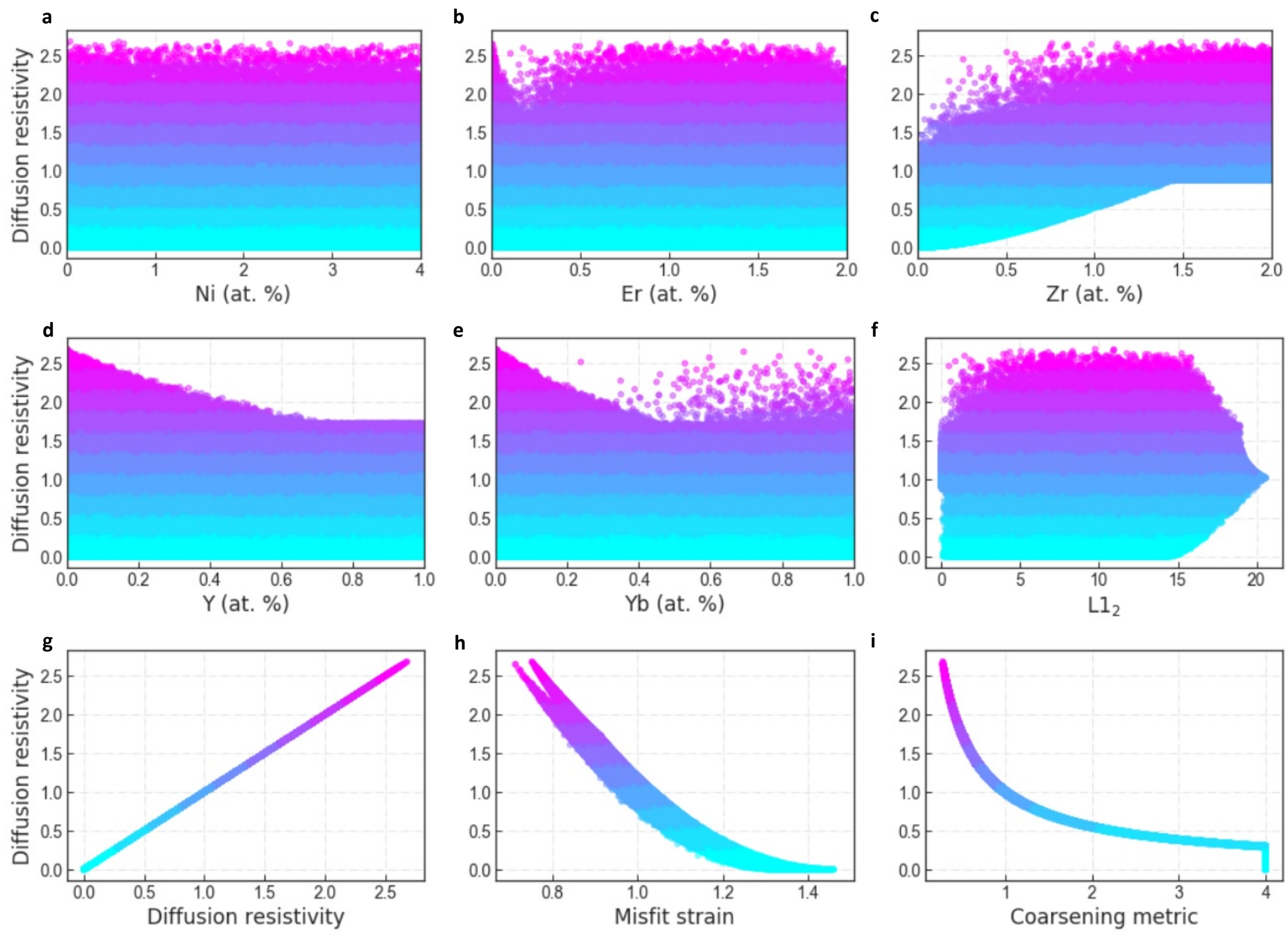

Fig. S5. Distribution of diffusion resistivity with respect to studied parameters. Increasing Zr has the greatest influence on increasing diffusion resistivity.

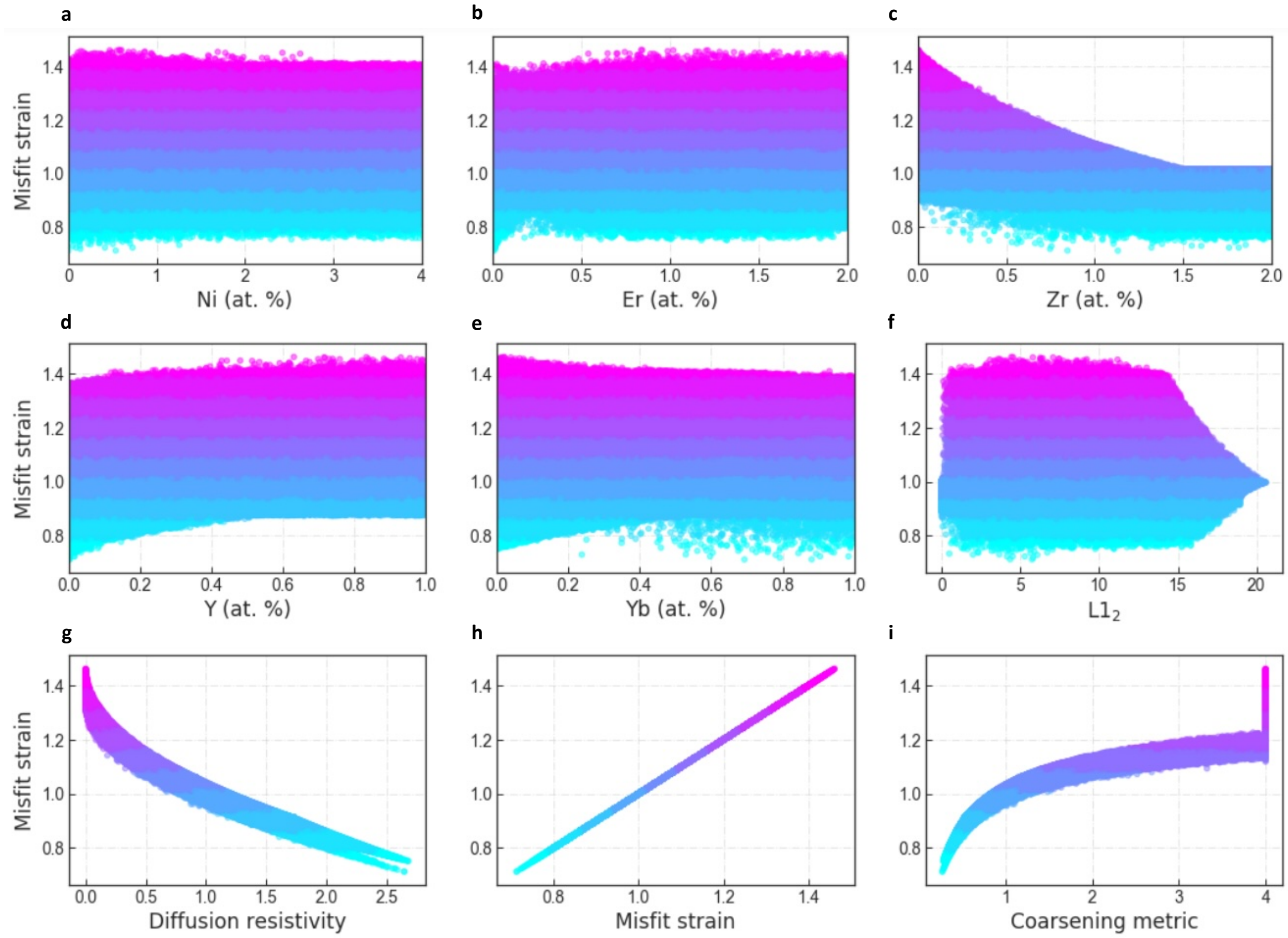

Fig. S6. Distribution of misfit strain (ε) with respect to studied parameters. Low coarsening metrics have low misfit strain.

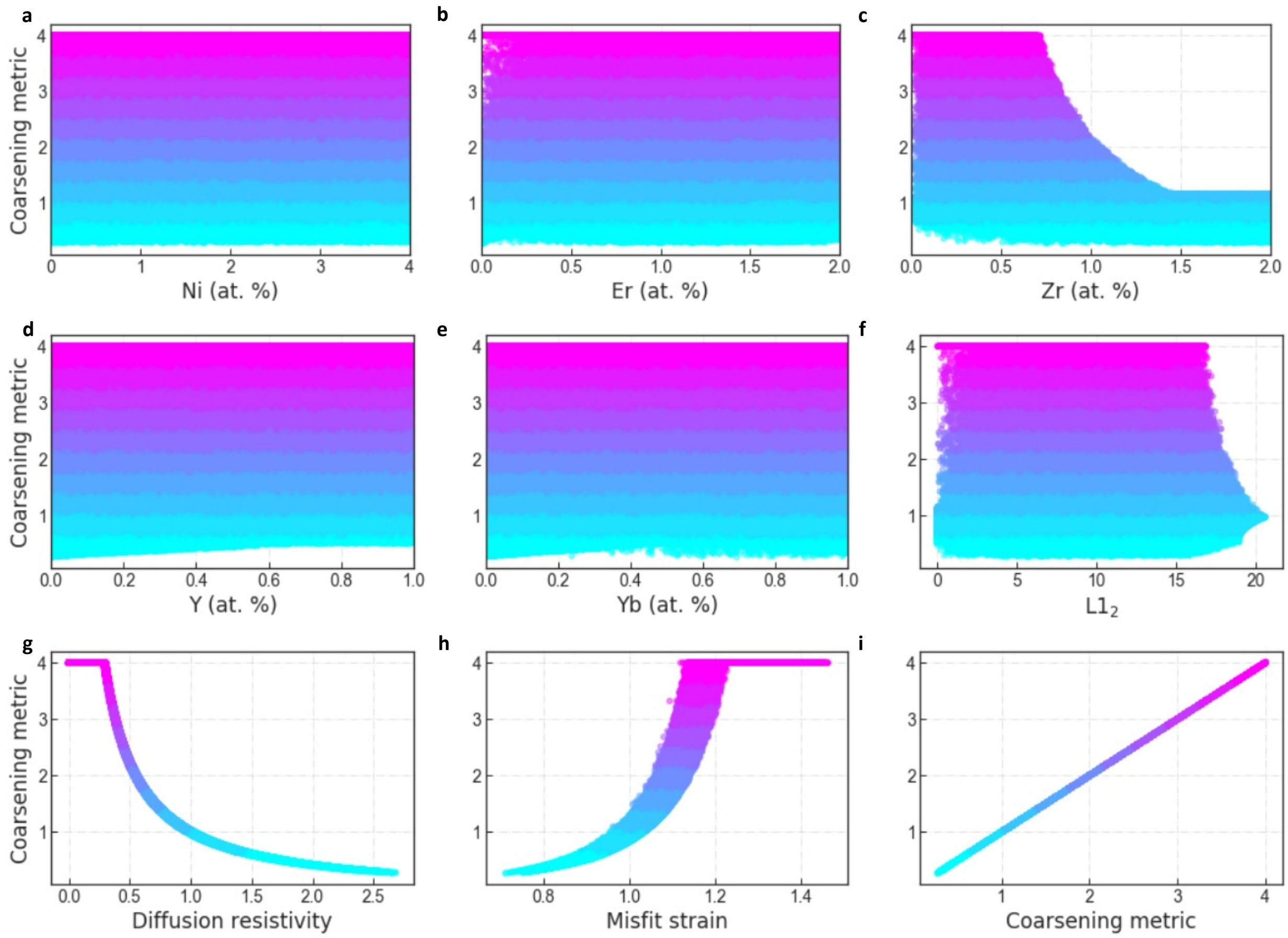

Fig. S7. Distribution of coarsening metric with respect to studied parameters. Increasing Zr has the most significant contribution to reduce the coarsening metric. Diffusion resistivity reveals a greater influence on coarsening metric than misfit strain (ε).

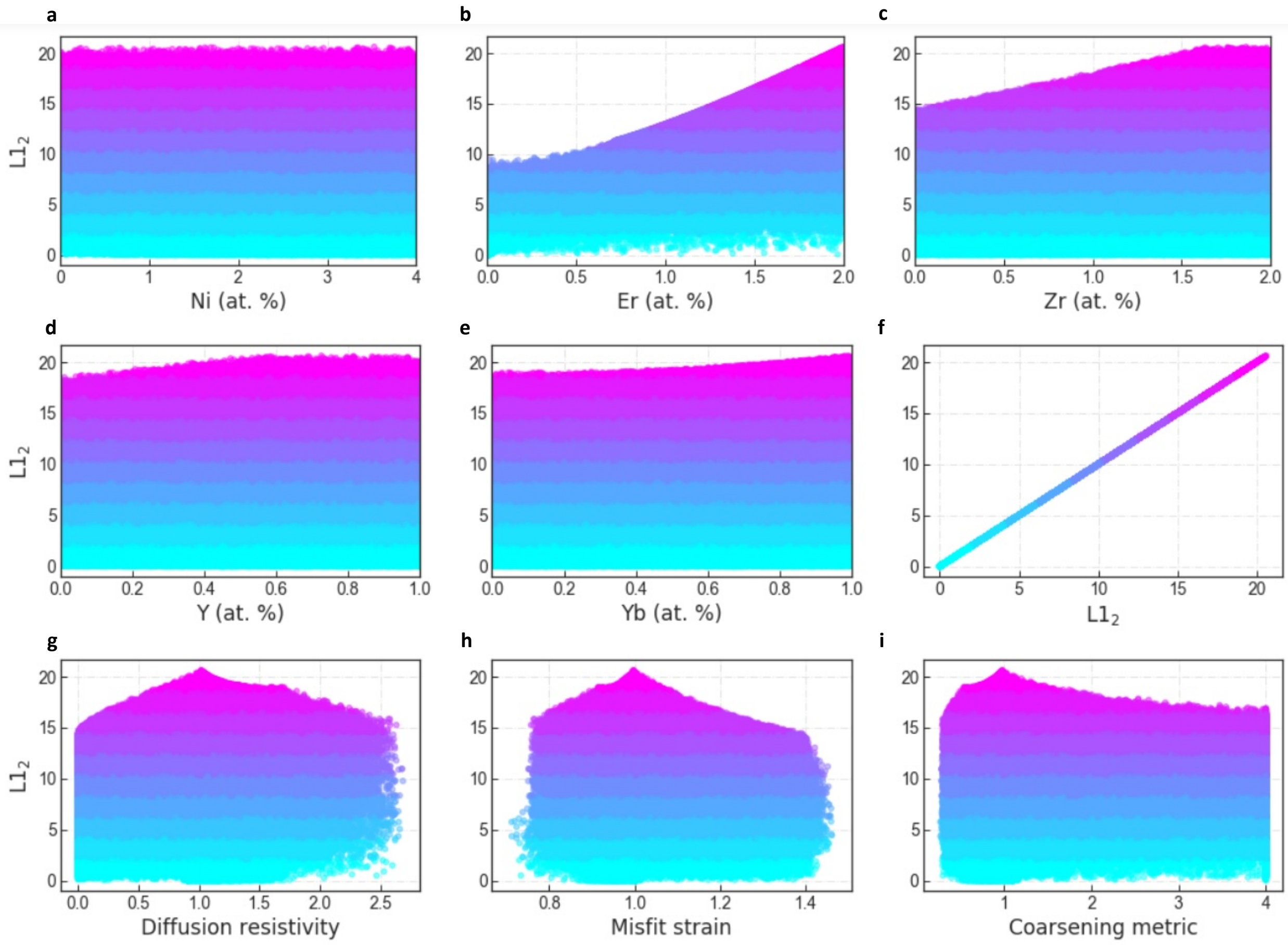

Fig. S8. Distribution of $L1_2$ at. % with respect to studied parameters. Increasing Er has the greatest influence on the $L1_2$ at. %.

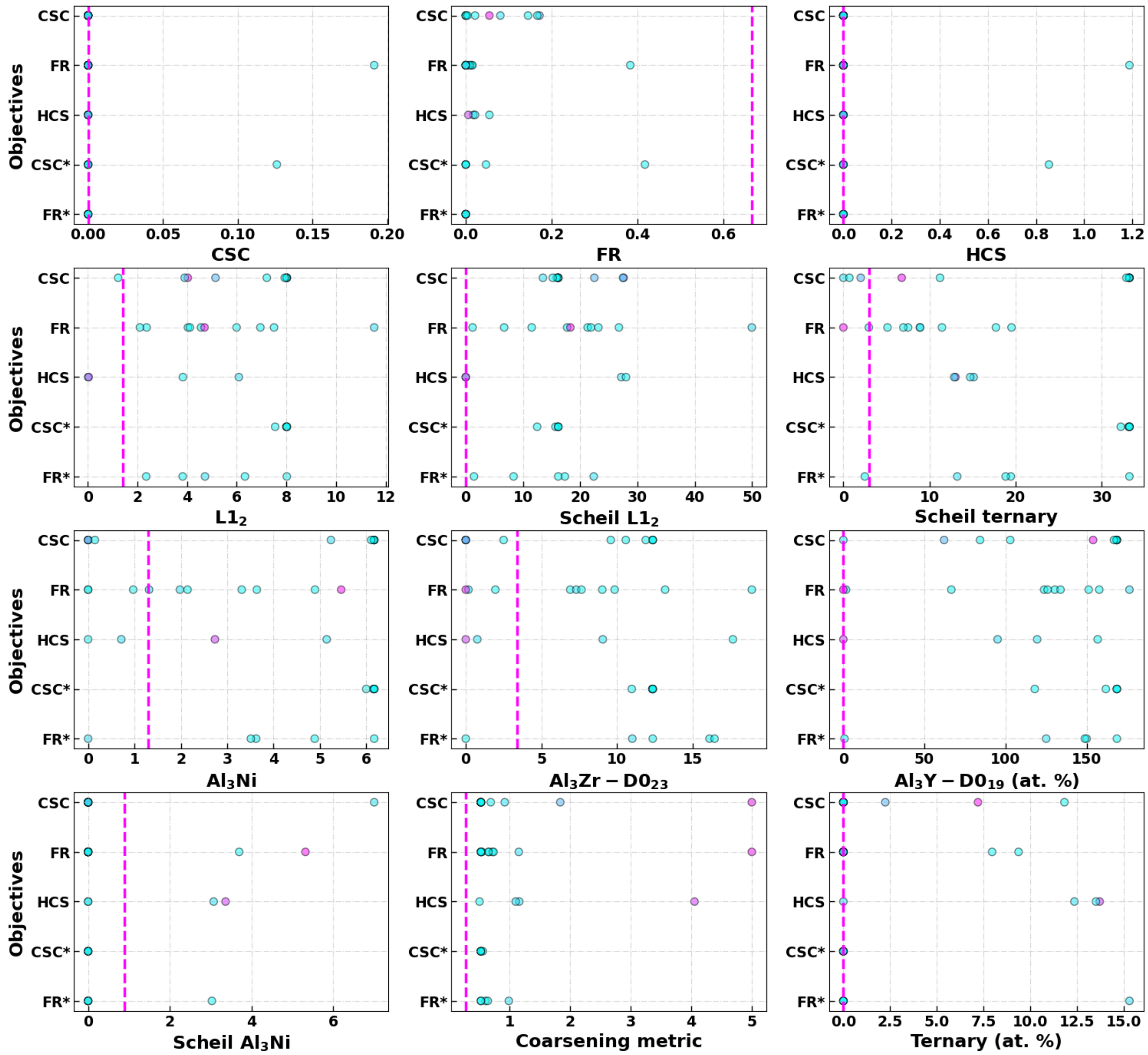

Fig. S9. Distribution of microstructural features/material descriptors for different design scenarios. The optimal value obtained via BO using objective functions constructed with only printability parameters (CSC, FR, and HCS) or combinations of these parameters with performance metrics (FR* & CSC*). Data were normalized against corresponding values in the benchmark design. CSC, HCS, Ternary (at. %), and $Al_3Y$ $D0_{19}$ (at. %) were not normalized as they were found to be zero in the benchmark design. ML-optimized design is plotted for each distribution as a vertical dashed line. Points are colored with coarsening metric, with blue being low and pink being high.

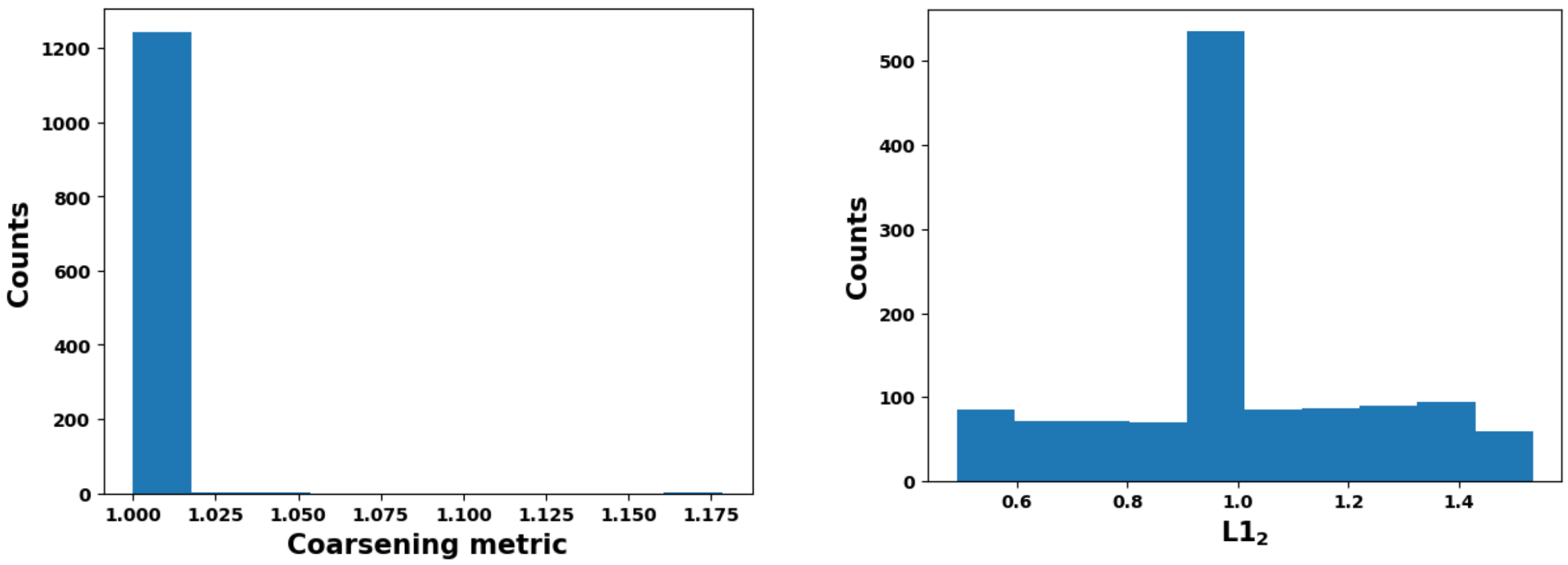

Fig. S10. Histogram of the coarsening metric and L1$_2$ at. % of 1250 designs varied from the ML-optimized design by up to 50% change in composition, normalized against the ML-optimized design. Both descriptors are highly stable under these variations.

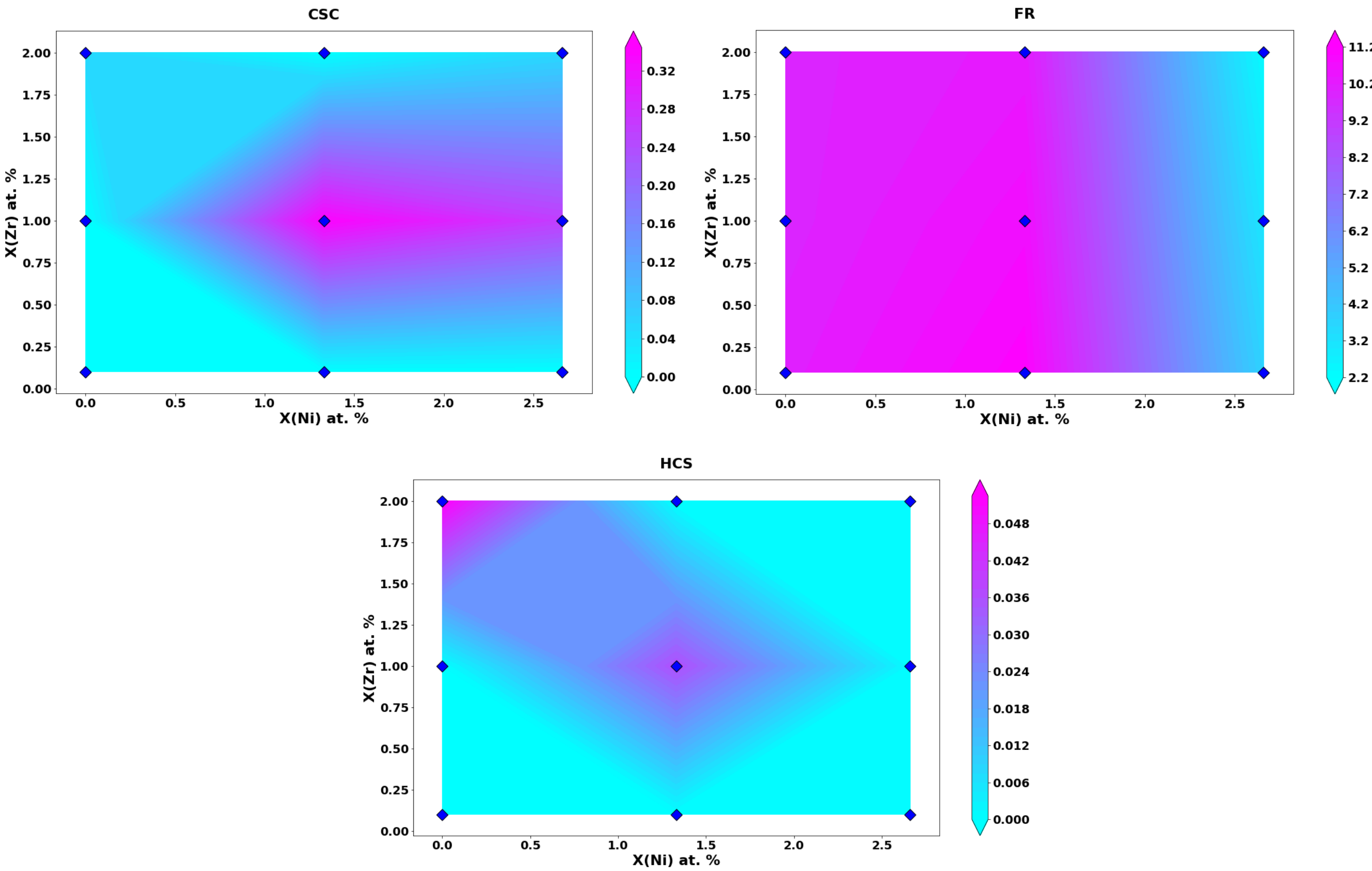

Fig. S11. Contour plot of freezing range (FR), cracking susceptibility coefficient (CSC), and hot cracking susceptibility (HCS) for various Ni and Zr at. %.

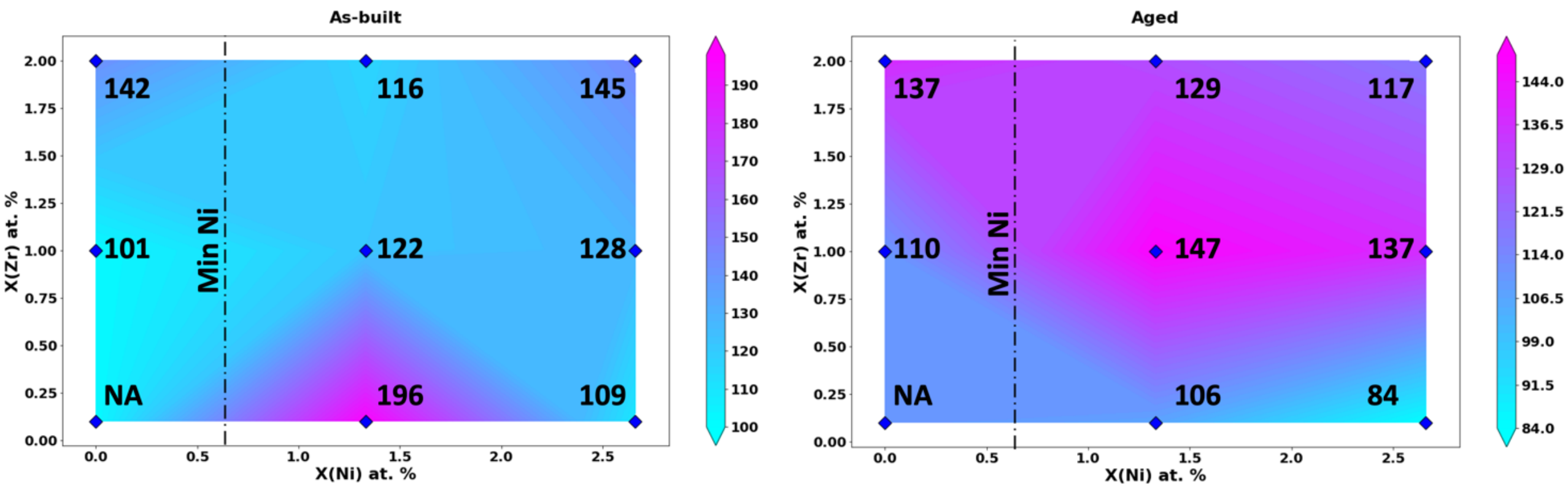

Fig. S12. Results of Vickers hardness testing on as-built and aged samples with different Ni and Zr contents. In both figures, the dashed lines define the min Ni At. % to have maximum ternary phase.

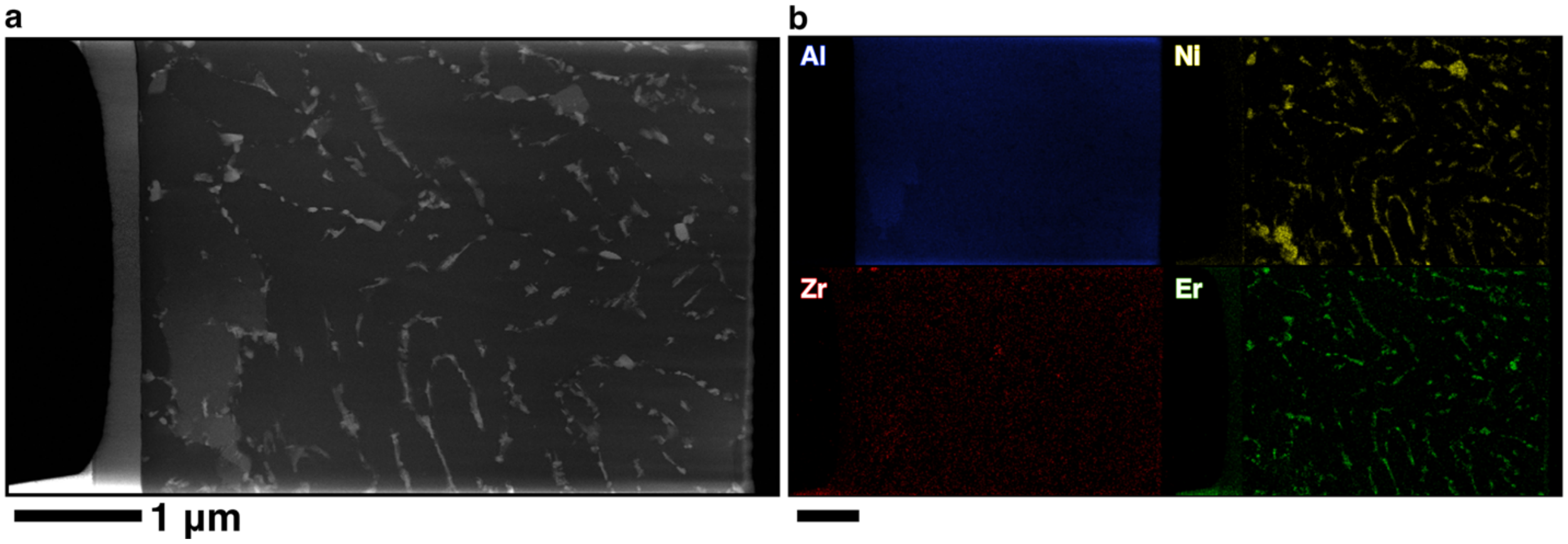

Fig. S13. a) Low magnification HAADF STEM image and b) EDS maps showing distributions of Al, Ni, Zr, and Er in the microstructure.

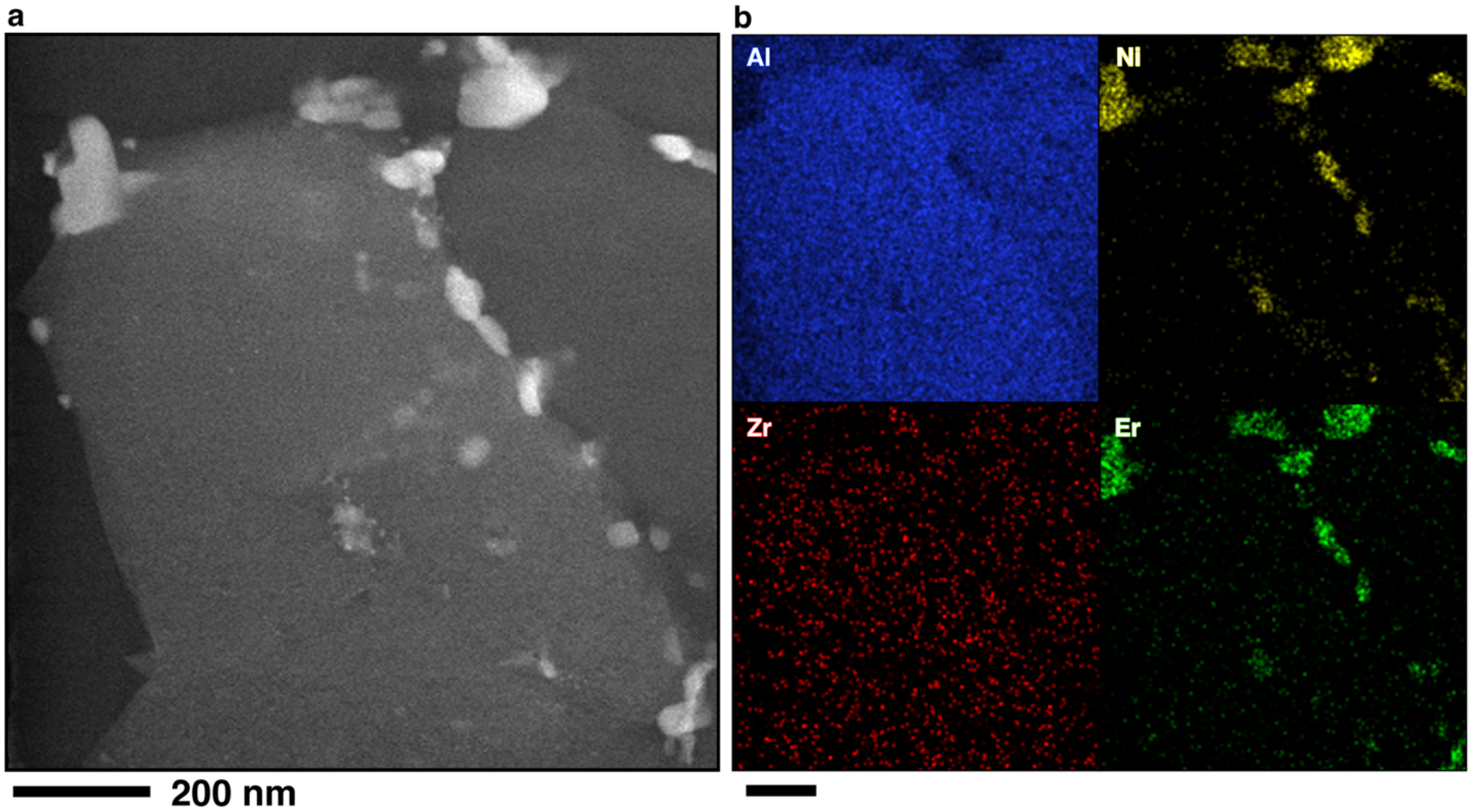

Fig. S14. a) HAADF STEM overview of the imaged grain along a <110>-type zone and b) corresponding EDS elemental maps showing distributions of Al, Ni, Zr, and Er.

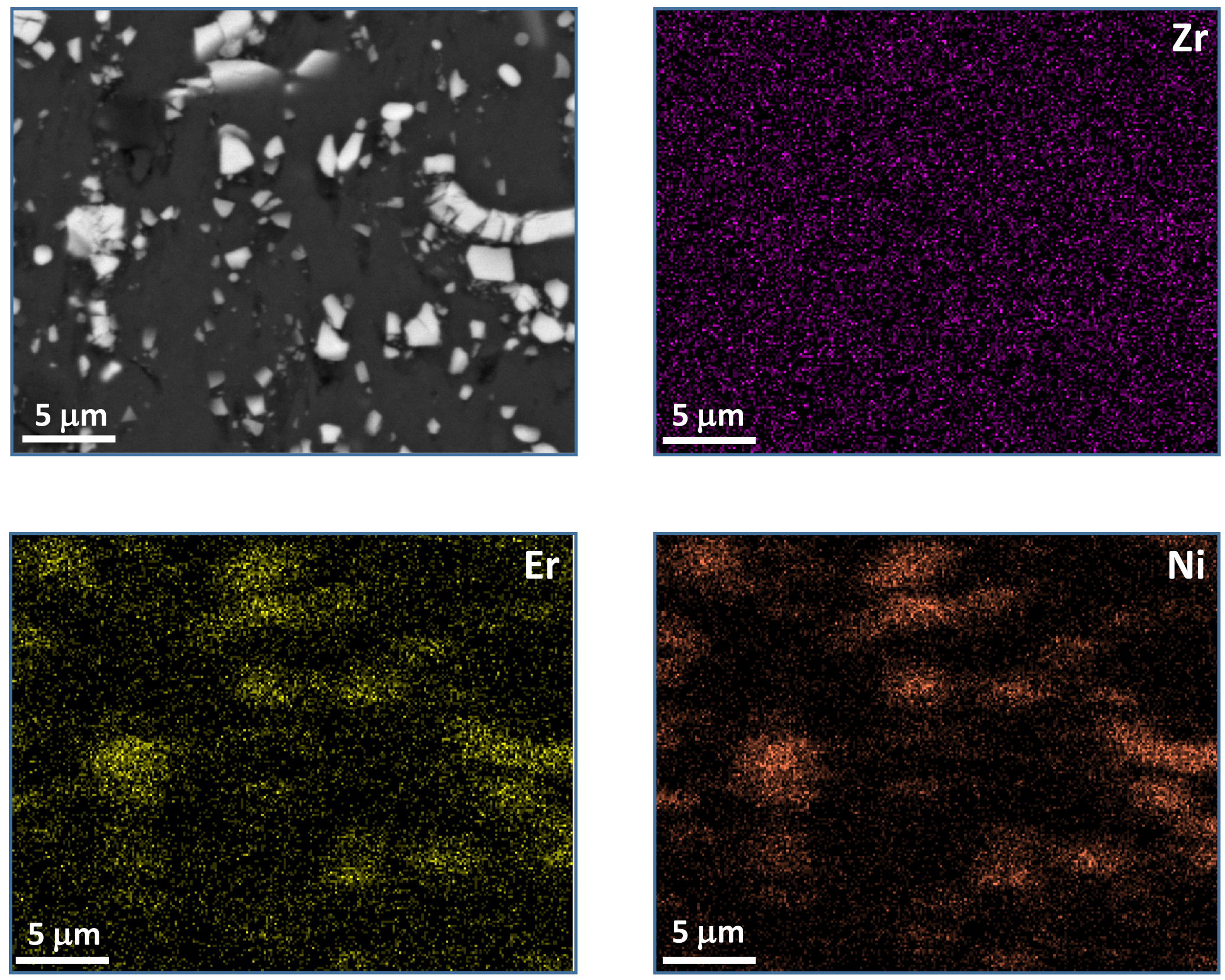

Fig. S15. SEM-EDS images of induction-melted sample from the optimized composition. Micro scale phases which are rich in Er and Ni contents are seen in these images.

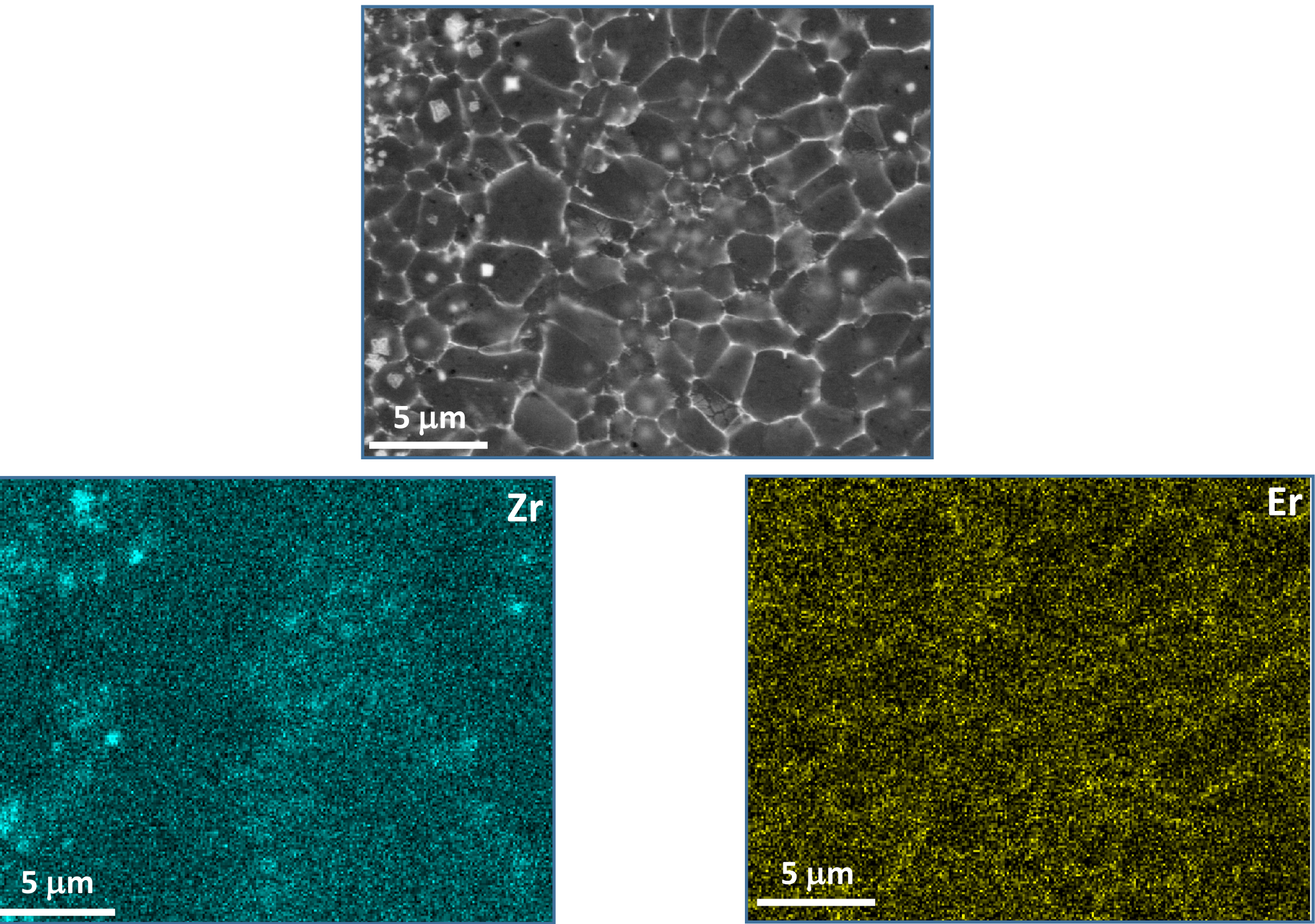

Fig. S16. SEM-EDS image of the sample with Er = 0.4 at. %, Zr = 1 at. %, and Ni = 0 at. %. Micro scale Zr rich areas are seen in some grains of this sample.

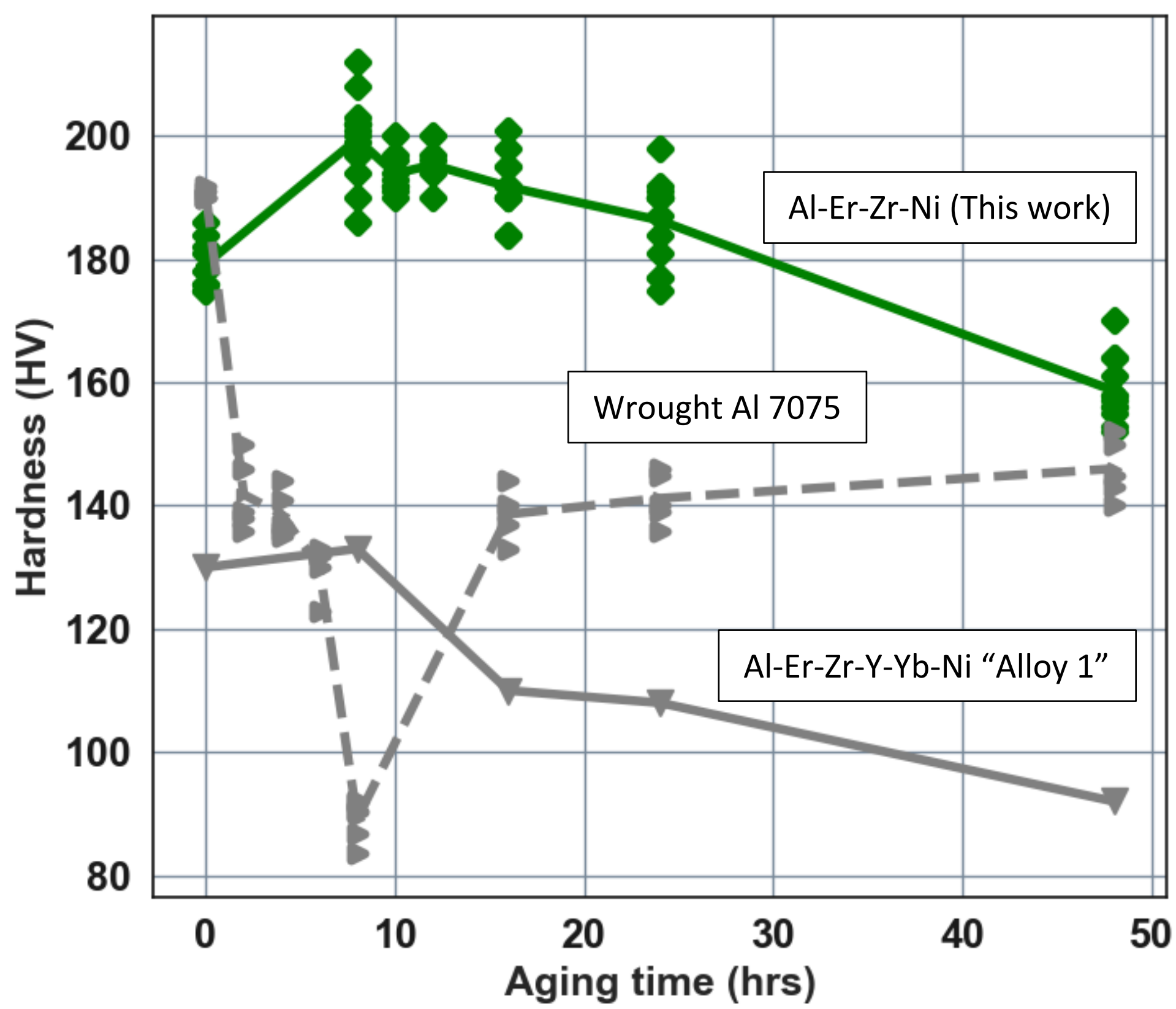

Fig. S17. Distribution of the hardness for experimented samples showed in Fig. 1.

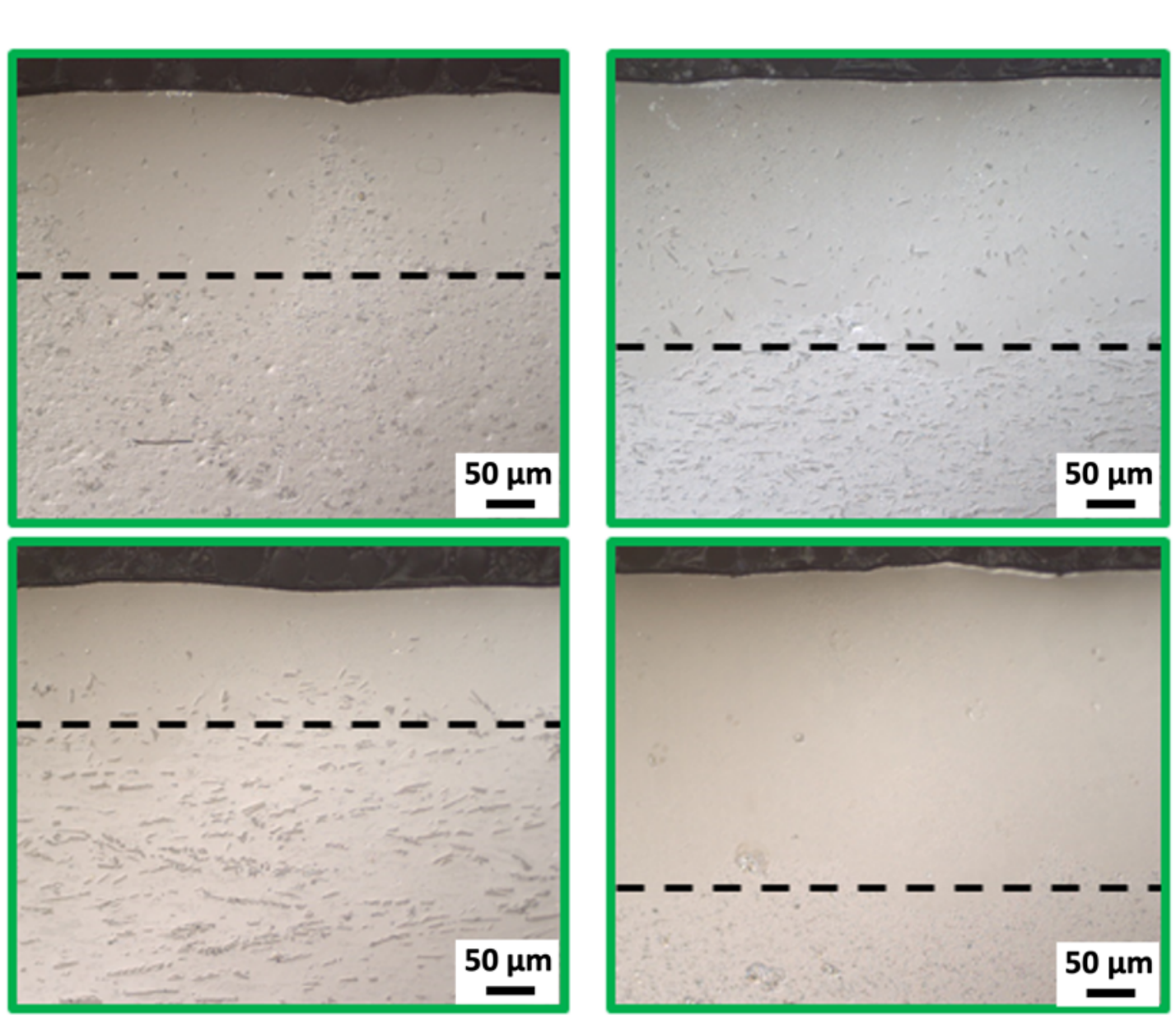

Fig. S18. Induction melted samples after multiple laser path scanning. No cracks were detected in melted areas (above dashed lines). These four samples are the four left down diamonds in Supplementary Fig. S11.

| Regression technique | Hyperparameters | Optimum hyperparameters |
| --- | --- | --- |
| Neural network (NN) | # of Conv layers = 1, 2, 3; Conv features = 64, 128, 256; # of FC layers = 2, 3, 4; FC neuron # = 1, 32, 32, 64, 64; Batch size = 32, 64, 128, 256, 512, 1024, 2048; Training loop number = 100, 1000, 10,000, 150,000, 200,000, 300,000; Learning rate = $10^{-4}$, $10^{-3}$, and $10^{-2}$; Regularization factor $L_1$ = 0, $10^{-6}$, $10^{-5}$; Regularization factor $L_2$ = 0, $10^{-2}$, $10^{-3}$ | # of Conv layers = 2; Conv features = 64; FC layers = 2; FC neuron # = 32, 1; Batch size = 128; Training loop number = 300,000; Learning rate = $10^{-3}$; $L_1$ = 0; $L_2$ = 0 |
| K-nearest-neighbors (KNN) | n_neighbors = 1, 2, 3, 4, 5, 6, 7, 8 | n_neighbors = 7 |
| Support vector machine (SVM) | kernel = 'rbf'; C = 1, 10, 100, 1000 | C = 1000 |
| Random forest (RF) | Max_depth = 2, 4, 6, 8, 10; n_estimators = 20, 40, 60, 80, 100, 120 | Max_depth = 10; n_estimators = 60 |
| Extreme gradient boost (XGB) | Max_depth = 2, 4, 6, 8, 10; num_boost_round = 20, 40, 60, 80, 100, 120 | Max_depth = 10; num_boost_round = 20 |
| Linear regression (LR) | | |

Table S1. Hyperparameters which were parametric studied for each regressor to develop the manifold.

| Composition (Wt. fraction) | Al | Er | Zr | Ni |
|---|---|---|---|---|
| Matrix | 0.9947 | 8.2e-05 | 0.009272 | 0.000543 |
| $L1_2$ | 0.79776 | 0.006912 | 0.193834 | 0.000475 |
| Ternary | 0.68717 | 0.91445 | 0.079137 | 0.138222 |
| Bulk | 0.9218 | 0.0237 | 0.029 | 0.0255 |

Table S2. Measured compositions for the matrix, $L1_2$, and ternary phases.